\documentclass[ %
 reprint,
superscriptaddress,
 amsmath,amssymb,
 aps,
floatfix,
]{revtex4-2}

\usepackage{graphicx}
\usepackage{dcolumn}
\usepackage{bm}
\usepackage{hyperref}
\usepackage[mathlines]{lineno}
\usepackage{xcolor, soul} 

\graphicspath{{ }}
\usepackage{tabularx}
\usepackage{adjustbox}
\usepackage{multirow}
\usepackage{caption}
\usepackage{physics}
\hypersetup{breaklinks=true}

\begin{document}

\preprint{APS/123-QED}

\title[On the detectability of higher harmonics with LISA]{On the detectability of higher harmonics with LISA}

\author{Chantal Pitte}
    \email{chantal.pitte@cea.fr}
\affiliation{IRFU, CEA, Universit\'{e} Paris-Saclay, F-91191, Gif-sur-Yvette, France.}

\author{Quentin Baghi}
\affiliation{IRFU, CEA, Universit\'{e} Paris-Saclay, F-91191, Gif-sur-Yvette, France.}

\author{Sylvain Marsat}
\affiliation{Laboratoire des 2 Infinis - Toulouse (L2IT-IN2P3), Universit\'{e} de Toulouse, CNRS, UPS, F-31062 Toulouse Cedex 9, France.}

\author{Marc Besançon}
\affiliation{IRFU, CEA, Universit\'{e} Paris-Saclay, F-91191, Gif-sur-Yvette, France.}

\author{Antoine Petiteau}
\affiliation{IRFU, CEA, Universit\'{e} Paris-Saclay, F-91191, Gif-sur-Yvette, France.}
\affiliation{Astroparticule et Cosmologie, Universit{\'e} de Paris, CNRS, Paris, 75013, France}

\date{\today}

\begin{abstract}
Supermassive black hole binaries (SMBHBs) are expected to be detected by the future space-based gravitational-wave detector LISA with a large signal-to-noise ratio (SNR). This prospect enhances the possibility of differentiating higher harmonics in the inspiral-merger-ringdown (IMR) waveform. In this study, we test the ability of LISA to identify the presence of different modes in the IMR waveform from a SMBHB. We analyze the contribution of each mode to the total SNR for different sources. Higher modes, in particular the mode $(3, 3)$ and $(4, 4)$, can dominate the signal observed through the LISA detector for SMBHB of the order of $10^8 M_\odot$. With Bayesian analysis, we can discriminate models with different harmonics. While spherical harmonics are often considered orthogonal, we observe it is not the case in the merger-ringdown phase observed by LISA. Omitting harmonics not only diminishes the SNR but can also lead to biased parameter estimates. We analyze the bias for each model in a source example and quantify the threshold SNR where we can expect the parameter bias to be comparable to the statistical error. By computing the waveform model error with the Fisher approximation and comparing it with the posterior distribution from our sampler results, we can evaluate the veracity of the analytical bias, which converges with the sampler results as more harmonics are introduced. To conclude, SMBHB events with SNR of a few hundred, as expected in LISA, are required to use templates with at least modes $(2, 2)$, $(2, 1)$, $(3, 3)$, $(3, 2)$, $(4, 4)$, $(4, 3)$ to estimate all intrinsic parameters correctly. Our work highlights the importance of higher modes to describe the gravitational waveform of events detected by LISA.
\end{abstract}


\maketitle

\section{\label{sec1}Introduction}

In the next decade, the Laser Interferometer Space Antenna (LISA) \cite{LISA_Proposal2017,SciRD} will leave Earth on the quest to find new gravitational-wave (GW) sources. The most intense signals are expected to result from the coalescence of supermassive black hole binaries (SMBHBs). The predicted rate depends on the population and evolution model~\cite{white_astro} and vary from 1 to 100 per year with a signal-to-noise ratio (SNR) ranging from a few tens up to thousands~\cite{Sesana_2021}. These high-SNR sources allow testing general relativity (GR) with black hole binaries (BHBs). One way to put Einstein's theory to the test with black hole binary coalescences is to probe the no-hair theorem \cite{Carter_26.331, Detweiler_1977} through the study of quasinormal modes (QNMs)~\cite{Dreyer_2004}.

After two compact objects have merged into one, a final black hole (BH) in a perturbed state is expected to be obtained. As it stabilizes to quiescence during the ringdown regime, it will emit gravitational radiation that can be described as a superposition of sinusoidal oscillations decaying over, i.e., the QNMs \cite{1973ApJ_P, chandra, Nollert1999QuasinormalMT, Kokkotas_1999}. Each QNM has an associated complex frequency labeled by $(l,m,n)$, including polar, azimuthal, and overtone indices. In GR, these solutions are entirely determined by the final black hole's mass and spin $(M_f, a_f)$ \cite{Echevarria_40.3194, Berti_2006}. Nonetheless, the corresponding amplitude and phase of each oscillation depend on the characteristics of the progenitors (the initial BHs) and their relative orientation towards the observer \cite{Kamaretsos_2012, Kamaretsos_2012_2}.

The spin-weighted `spheroidal' harmonics are the eigenfunctions of the ringdown. They can be projected into spin-weighted spherical harmonics, which are the basis of numerical relativity (NR) (see Eq.~(3.7) in \cite{1973ApJ_P}, also see \cite{Berti_2014, London_2014, Kelly_2013} for further information and mixing of the modes. This representation is conventionally used to describe the full inspiral-merger-ringdown (IMR) waveform, as it is done, for instance, in phenomenological models, where only the fundamental overtone (n=0) is included in each spherical harmonic in the ringdown regime. From now on, we will refer to spin-weighted spherical harmonics components of the IMR waveform as \textit{modes} and will index them by their mode numbers $(l,m)$.
The response of LISA can be included as a transfer function in the frequency domain separately for each mode, 
allowing us to produce fast phenomenological IMR waveforms \cite{PhenomD, PhenomDII, London_2018} with the appropriate instrumental response \cite{Marsat_18}.

Investigating the role of higher harmonics in SMBHB signals is essential because of the high SNR these events will have in LISA. The analysis of such strong signals will be sensitive to many subdominant features in the waveform and, in particular, to higher harmonics beyond the dominant $(2,2)$ mode. Modes with different $m$ are often considered orthogonal since their phases scale differently with the orbital phase as $m \phi_{\rm orb}$, leading to destructive interference. In LISA, this is no longer the case for the merger-ringdown phase, where a large SNR is accumulated over only a few wave cycles. Cross-terms of the harmonics yield an SNR contribution in the likelihood, which can also affect the parameter inference. Thus, the absence of higher harmonics in the template will induce biases in the parameter estimation.

The correct estimation of the parameters of the full waveform is crucial, in particular, to test the no-hair theorem. There are two main approaches to probing the no-hair hypothesis. The first one uses two QNM parameters to find the ``hairs" of the final BH, i.e., final mass and spin $(M_f, a_f)$, while a third QNM parameter, or preferably more, are used to check for consistency~\cite{Dreyer_2004, Gossan_2012}. The second approach compares the estimated final parameters from the full IMR or pre-merger waveform with those obtained from only the ringdown~\cite{PhysRevLett.116.221101}. Their inequality can be understood as a possible deviation from GR. Note that in the latter case, the measure of only one harmonic is necessary to test the hypothesis. A third method called the ``merger-ringdown" test has recently been proposed to check for consistency between both regimes. In the same line as the previous method, it is based on the relation of the amplitude and the phase of the QNMs with the properties of the progenitors~\cite{Bhagwat_2021}. 
To test the hypothesis with any method, one should know the intrinsic parameters with enough accuracy and precision. They include the total mass $M$, the mass ratio $q$, and the individual spins $S_i$ of the initial BHs. With the progenitors' parameters, one can compute the values of final mass and spin $(M_f, a_f)$ (see, e.g., \cite{Pan_2011, PhenomD}) thus allowing one to perform BH spectroscopy, i.e., the study of the BH harmonics. Nonetheless, one should also determine the extrinsic parameters from the IMR or pre-merger inference to analyze the ringdown regime.

Several analyses of black hole spectroscopy have been proposed over the years \cite{Berti_2006, Berti_2009,  Baibhav_2019, Baibhav_2020, Giesler_2019, Bhagwat_2020}. The work from \cite{Gossan_2012} was the first to use Bayesian analysis to study deviations from GR in the context of the first method; but also see \cite{Thrane_2017, Meidam_2014}. Concerning the second approach, studies of the detectability of overtones ($n \geq 1 $) or higher angular modes ($l,m \geq 2 $) have been performed by the LIGO-Virgo-Kagra (LVK) collaboration as well as by other authors \cite{PhysRevLett.116.221101, PhysRevD.103.122002, Isi_2019, Bustillo_2021, Carullo_2019, Capano_2020, Cotesta_2022}. Tension between different authors is observed regarding the presence of higher modes for the detected event GW150914~\cite{Abbott_2016}. We recommend however to read \cite{baibhav2023agnostic} for a thorough analysis on the detectability of the overtones. The effective-one-body (EOB) formalism \cite{Buonanno_EOB, Buonanno_EOB2, Bohe_SEOB} (full IMR waveform for spinning or non-spinning binaries) was also adapted to study possible deviations from GR, namely pSEOBNR and pEOBNR \cite{ maggio2023tests, Brito_2018}, for events in the LVK frequency range. To the best of our knowledge, the full LISA response (including high-frequency effects) has not been taken into account to study ringdown signals. 

In this study, we evaluate our ability to identify and differentiate modes of a plausible source detected by LISA and investigate the possible consequences of ignoring modes. To this end, we make use of the software \texttt{lisabeta}~\cite{Sylvain_git}, which incorporates LISA's response to the source waveform, as described in Section~\ref{sec2}. We continue explaining the study methodology in Section~\ref{sec3}. In Section~\ref{sec4}, we analyze the contribution of the modes to the total SNR for general cases. Then we focus on the impact of mode contributions on estimating parameters for a specific event in Section~\ref{sec5}. We also analyze the errors from using an incorrect template relative to the source SNR. Finally, we summarize our conclusions in Section~\ref{sec6}.


\section{Supermassive black hole waveforms in LISA}\label{sec2}

To study signals observed by LISA, we must incorporate its instrumental response to the GW signature produced by an event. We use the \texttt{lisabeta} software developed for this purpose, which accounts for several instrumental effects in LISA. We review the main features implemented in \texttt{lisabeta} that are particularly interesting to our study in the following subsections. More detailed information can be found in \cite{Marsat_18}.

\subsection{Waveform in the source and detector frames}

We generate the SMBHB waveform with PhenomHM \cite{London_2018}, a phenomenological approach based on PhenomD \cite{PhenomD, PhenomDII} for non-precessing binaries. In addition to the dominant quadrupole $(l=2, m=2)$, higher modes are introduced, including $(l,m) =$ $(2, 1)$, $(3, 3)$, $(3, 2)$, $(4, 4)$, $(4, 3)$. Incorporating these higher harmonics brings crucial complementary information, not only to the ringdown part of the signal but also to the inspiral and merger, which has been shown to narrow down the posterior of extrinsic parameters in the inference \cite{Arun_2007,Arun_2007b, Trias_2008, Arun_2009, Marsat_2021, veccio}.

The gravitational-wave signal can be decomposed in spin-weighted spherical harmonics  ${}_{-2}Y_{lm}$, which depend on the orientation of the emission parametrized by the inclination $\iota$ and the phase $\varphi$. The polarizations $h_{+}$ and $h_{\times}$ of a GW are related to their harmonics by
\begin{equation}
    h_{+} - i h_{\times} = \sum_{l \geq 2} \sum_{m = -l}^{l} {}_{-2}Y_{lm}(\iota, \varphi) h_{lm},
\end{equation}
where each mode can be described in terms of an amplitude $A_{lm}$ and a phase $\Phi_{lm}$ that depends on the intrinsic parameters of the source 
\begin{equation}
    h_{lm} = A_{lm} \, e^{-i \Phi_{lm}}.
\end{equation}
In the frequency domain, non-precessing binary systems have an advantageous symmetry relation between progrades and retrogrades modes ($m$ and $-m$). It allows us to describe each polarization as
\begin{equation}{\label{eq_Klm}}
    \tilde{h}_{+, \times}(f) = \sum_{l} \sum_{m>0} K_{lm}^{+,\times} \tilde{h}_{lm}(f),
\end{equation}
where we introduced 
\begin{subequations}
\begin{align}
    K_{lm}^+ = & \frac{1}{2} \left( {}_{-2}Y_{lm} + (-1)^l {}_{-2}Y_{l-m}^*\right),\\
    K_{lm}^\times = & \frac{i}{2} \left( {}_{-2}Y_{lm} - (-1)^l {}_{-2}Y_{l-m}^*\right).
\end{align}
\end{subequations}

The GW strain in the traceless-transverse gauge is expressed as
\begin{equation}
    h^{TT} = \textbf{e}_+ h_{+} + \textbf{e}_\times h_{\times},
\end{equation}
where \textbf{e}$_{+,\times}$ are the polarization tensors,
\begin{subequations}\label{eq_tensor}
\begin{align}
    \textbf{e}_{+} = & \textbf{u} \otimes \textbf{u} - \textbf{v} \otimes \textbf{v}, \\
    \textbf{e}_{\times} = & \textbf{u} \otimes \textbf{v}  + \textbf{v} \otimes \textbf{u}.
\end{align}
\end{subequations}
Vectors \textbf{v} and \textbf{u} together with the propagation vector \textbf{k} in spherical coordinates locate the source in the observational frame,
\begin{subequations}\label{eq_vector}
\begin{align}
    \textbf{u} = & \{\sin \lambda, \cos \lambda, 0 \},\\
    \textbf{v} = & \{-\sin \beta \cos \lambda, -\sin \beta \sin \lambda, \cos \beta \},\\
    \textbf{k} = & \{-\cos \beta \cos \lambda, -\cos \beta \sin \lambda, -\sin \beta \}.
\end{align}
\end{subequations}
with $(\beta, \lambda)$ as the ecliptic latitude and longitude.

Combining previous equations, we obtain the final expression

\begin{equation}\label{ssb_strain}
    \tilde{h}^{TT}(f) = \sum_{l,m} P_{lm} \tilde{h}_{lm}(f),
\end{equation}
where 
\begin{align}\label{eq:pol}
\begin{split}
    P_{lm}  & = \textbf{e}_+ K_{lm}^+ + \textbf{e}_\times K_{lm}^\times\\
     & = \frac{1}{2} \left[ {}_{-2}Y_{lm} (\textbf{e}_+ + i\, \textbf{e}_\times) e^{-2i\psi} \right.\\ 
     & \qquad \left. + (-1)^l {}_{-2}Y_{l-m}^* (\textbf{e}_+ - i\, \textbf{e}_\times) e^{2i\psi}\right].
\end{split}
\end{align}
Note that $P_{lm}$ depends not only on $(l,m)$ and $(\iota, \varphi)$ due to the spherical harmonics, but also on the parameters defining the reference frame, such as the sky localization~$(\beta, \lambda)$ and the polarization~$\psi$. 

\subsection{LISA response}

The LISA constellation comprises three spacecraft (S/C) deployed triangularly.  As a GW travels across one arm of the constellation, the detectors at both ends of the arm will read the frequency shift between the sent and received signals. This response is known as the link response. One advantage of this triangular setup is that it allows for forming multiple interferometers with different combinations of the links. Each link response is defined as 
\begingroup\makeatletter\def\f@size{9}\check@mathfonts
\def\maketag@@@#1{\hbox{\m@th\normalsize#1}}%
\begin{align}
\begin{split}
    y_{rs}(t_r) \simeq & \frac{1}{2 \left( 1 - \hat{\textbf{k}} \cdot \hat{\textbf{n}}_{rs} (t_r)\right)} \left[ 
    H_{rs}\left( t_r - L_{rs}(t_r) - \right. \right. \\  
    & \left. \left. \hat{\textbf{k}} \cdot \textbf{x}_r(t_r)\right)  - H_{rs} \left( t_r - \hat{\textbf{k}} \cdot \textbf{x}_s(t_r) \right) \right] ,
\end{split}\label{eq:response1}
\end{align}
\endgroup
where we used $c=1$. The position of the S/C is designated by \textbf{x}$_{r,s}$ where the sub-index $s$ stands for the sender, while $r$ stands for the receiver. Note the ordering of those indices, which follows the last convention adopted in \cite{Bayle_2021}. Their values go from 1 to 3, indexing the three S/C. Therefore six combinations are produced, resulting in six different links. The unit vector $\hat{\textbf{k}}$ defines the direction of propagation of the GW, while $\hat{\textbf{n}}_{rs}$ is the direction of the beam. Finally, $L_{rs}$ is the arm's length between the two S/C, and $H_{rs}$ is the source's gravitational strain projected into the arm, 

\begin{align}
\begin{split}
    H_{rs}(t) & = \, \left( h_{+} (t) \cos 2\psi \, - \right.  \\
    & \left. \hspace{8mm} h_{\times}(t) \sin 2\psi\right) \, \hat{\textbf{n}}_{rs}(t) \cdot \textbf{e}_{+} \cdot \hat{\textbf{n}}_{rs}(t) \\
    & + \, \left( h_{+}(t) \sin 2\psi \, + \right.  \\
    & \left. \hspace{8mm} h_{\times}(t) \cos 2\psi\right) \, \hat{\textbf{n}}_{rs}(t) \cdot\textbf{e}_{\times} \cdot  \hat{\textbf{n}}_{rs}(t).
\end{split}\label{eq:response2}
\end{align}

The constellation as a whole will follow Earth in the same orbit. However, each S/C will follow its own orbit around the sun and within the constellation. The orbital motion translates into a time variation in the orientation of the detector relative to the solar system barycenter frame (SSB). Note that this introduces modulations on the signal observed by LISA, as is noticeable from the explicit time-dependent prefactors and delays in the instrumental response given by Eqs.~\eqref{eq:response1} and~\eqref{eq:response2}. In LISA's frequency band, the observation of SMBHBs can last from days to months, depending on the total mass and frequency evolution. Therefore, their waveform can be strongly affected by these modulations. Additionally, in the post-processing, beams are combined in the time-delay interferometry (TDI, see Ref.~\cite{Tinto_1999, Vallisneri_2005} and Sec.~\ref{sec:tdi}) to cancel laser frequency noise. As a result, the time delays and their variations leave an imprint in the measured signal.

\subsubsection{TDI and frequency-domain formulation}\label{sec:tdi}

Time delay interferometry was first proposed by Armstrong, Estabrook, and Tinto \cite{Armstrong_1999} as a solution to cancel the dominant noise produced by fluctuations in the laser frequency. The idea is to combine the links linearly with an adequate time delay to eliminate laser frequency noise. Different combinations have been proposed depending on the characteristics of the constellation. We can find what is known as a first-generation Michelson interferometer (TDI 1.5) for a stationary unequal-arm constellation and a second-generation Michelson interferometry (TDI 2.0) for a rotating unequal- and flexing-arm constellation. Note that any actual constellation will have flexing arms, which prompted the development of TDI~2.0.

It was shown (see e.g.~\cite{Marsat_18}) that it is possible to approximately express the response of the links in terms of the harmonics in the frequency domain as
\begin{equation}\label{eq_transfer}
    \tilde{y}_{rs} (f) = \sum_{lm} \mathcal{T}_{rs}^{lm} (f)\, \tilde{h}_{lm} (f) ,
\end{equation}
where $\mathcal{T}_{rs}^{lm}(f)= G_{rs}^{lm}(f, t_f^{lm})$ is a kernel carrying information on the modulation and time-delay of the links response. It is defined as 
\begingroup\makeatletter\def\f@size{9}\check@mathfonts
\def\maketag@@@#1{\hbox{\m@th\normalsize#1}}%
\begin{align}\label{eq_kernel}
\begin{split}
    G^{lm}_{rs}(f, t) & = \frac{i \pi f L_{rs}}{2} \, \text{sinc} \Big[\pi f L_{rs} \left( 1-\hat{\textbf{k}} \cdot \hat{\textbf{n}}_{rs}(t)\right) \Big] \cdot \\
    & e^{i \pi f \left( L_{rs} + \hat{\textbf{k}} \cdot [\textbf{x}_r(t) +\textbf{x}_s(t)]\right)} \, \hat{\textbf{n}}_{rs}(t) \cdot P_{lm} \cdot \hat{\textbf{n}}_{rs}(t).  
\end{split}
\end{align}
\endgroup
The frequency-domain response can be obtained thanks to a generalized relation giving a time-to-frequency correspondence. This relation extends the one found within the stationary phase approximation (SPA) \cite{Cutler_1994, Klein_2013}, which applies only to the slowly evolving inspiral signal, to the merger-ringdown part of the signal. It reads~\cite{Marsat_18, Marsat_2021}

\begin{equation}\label{eq_spa}
     t_f^{lm} \equiv - \frac{1}{2 \pi} \frac{d \Phi^{lm}}{d f},
\end{equation}
where $\Phi^{lm}$ is the phase of each mode (l,m). 
%

To simplify the transfer function in the frequency domain, some assumptions were made: 
\begin{itemize}
    \item the constellation forms an equilateral triangle, then the arms will remain equal and constant;
    \item $L_{12} = L_{21}$, i.e., the arm length is similar in both directions since the relative motion of the beam relative to the S/C (also known as the pointing-ahead effect) is not taken into account. 
\end{itemize}

With these assumptions and after factoring out several terms, we can write the response in terms of channels $A, E, T$; the optimal linear combinations of Michelson variables~\cite{Prince_2002} that are approximately orthogonal relative to the noise. They read
\begin{subequations}\label{eq_channels}
\begin{align}
    \tilde{A} =  & \frac{i \sqrt{2} \sin(2\pi f L)}{e^{-2 i\pi f L}} \left[ (1 + z) (\tilde{y}_{13} + \tilde{y}_{31}) \right. - \notag\\
    & \left. \tilde{y}_{32} - z \tilde{y}_{23} - \tilde{y}_{12} - z \tilde{y}_{21} \right],\\
    \tilde{E} =  & \frac{i \sqrt{2} \sin(2\pi f L)}{\sqrt{3}e^{-2 i\pi f L}} \left[ (1 - z) (\tilde{y}_{31} + \tilde{y}_{13}) \right. + \notag \\
    & \left. (2+z) (\tilde{y}_{21} - \tilde{y}_{23}) + (1+2z) (\tilde{y}_{12} - \tilde{y}_{32}) \right], \\
    \tilde{T} = & \frac{4 \sin(\pi f L) \sin(2\pi f L)}{\sqrt{3} e^{-3 i\pi f L}}  \left[ \tilde{y}_{12} - \tilde{y}_{21} \right. + \notag \\
    & \left.   \tilde{y}_{23} - \tilde{y}_{32} + \tilde{y}_{31} - \tilde{y}_{13} \right],
\end{align}
\end{subequations}

with $z\equiv e^{2i \pi f L}$.  

\section{Methodology}\label{sec3}

This study aims to quantify our ability to identify the presence of different modes in a SMBHB event detected by LISA. We expect the detectability of the modes to be related to the SNR itself and the relative SNR of each mode. In this framework, there are two main parameters to consider in the computation of the SNR: the luminosity distance and the mass. While the distance is just a scaling factor that affects all modes similarly, the mass (and therefore the frequency) moves the signal to lower and higher frequencies affecting the relative weight of the inspiral and merger-ringdown. Since the observed signal is characterized by the redshifted mass, we will distinguish between the source-frame mass and the observed (redshifted) mass.

As an analogy to the luminosity peak in the time domain, we use here an ad-hoc frequency-domain definition of \textit{frequency peak} as the frequency at the maximum value of the observed re-scaled TDI variable for each mode; see Eq.~(29) of~\cite{Marsat_2021} for a definition of re-scaled TDI variables. With the time-to-frequency correspondence in Eq.~\eqref{eq_spa}, this parameter lies about $2.4 t_M$ ($t_M = t \, c^3 / M \, G$, adimensional value) after the time of coalescence. In the waveform observed by LISA, this frequency peak lies in the ringdown regime, even though there is no clear starting point for it (see ongoing discussion on this topic~\cite{Berti_2007, London_2014, Bhagwat_2018, Bhagwat_2020, baibhav2023agnostic}).
The mass of the source will move the frequency peaks through the spectrum so that the contribution of each mode depends on the sensitivity of LISA at the frequency peaks. For this reason, we study the contribution of each mode in terms of the mass. Results are shown in Section~\ref{sec4}.

Once we know the contribution of different modes in the general case, we can focus on a specific event as an example. The method we use to assess the detectability of modes is Bayesian model comparison and parameter estimation. Each model corresponds to a different combination of modes in generating the waveform. This method allows us to compare different models based on Bayes factors, as we shall see. 

\subsection{Bayesian analysis}\label{sec_Bayes}

In Bayesian analysis, the posterior distribution of the strain parameters $\boldsymbol{\theta}$ given the observed data $d$, is expressed as
\begin{equation}\label{posterior}
    p(\boldsymbol{\theta} \vert d, M) = \frac{p(d\vert \boldsymbol{\theta}, M) \,  p(\boldsymbol{\theta} \vert M)}{p(d \vert M)},
\end{equation}
where $\boldsymbol{\theta}$ are the physical parameters of the source, $M$ is the model (and any other context) considered, $p(d \vert M)$ is the evidence, $p(\boldsymbol{\theta} \vert M)$ the prior of the parameters (in practice independent of $M$), and $p(d\vert \boldsymbol{\theta}, M)$ the likelihood. The three latter are also usually denoted by $\mathcal{Z}$,  $\pi(\theta)$, and $\mathcal{L}(\theta)$, as the evidence, the prior, and the likelihood, respectively. In the following, we drop the indication of the model $M$ in the equations.

The likelihood for a Gaussian noise with a covariance matrix $\mathbf{C}$ takes the form
\begin{equation}\label{eq:log}
    \mathcal{L} =  \frac{1}{\sqrt{\text{det}(2\pi \mathbf{C})}}e^{-\frac{1}{2}(d -h(\boldsymbol{\theta}))^{\dagger}\mathbf{C}^{-1}(d -h(\boldsymbol{\theta}))}.
\end{equation}

From now on, we drop the tilde notation for simplicity, even though we work in the frequency domain. One advantage of working in the frequency domain is that each frequency bin is approximately independent when dealing with stationary noise. The noise covariance can therefore be diagonalized and represented by its power spectral density (PSD). 

We introduce the definition of the inner product as
\begin{equation}
    (a \vert b) = 4 \text{Re} \int_0^\infty  \frac{a(f) b^*(f)}{S_{n}(f)} df,
\end{equation}
where $S_{n}$ is the noise's PSD and $b^*$ is the complex conjugate or $b$. Eq.~\eqref{eq:log} can thus be rewritten as
\begin{equation}
    \ln \mathcal{L} = -\frac{1}{2}(d - h(\boldsymbol{\theta})\vert d -h(\boldsymbol{\theta})) + 
    \text{const}.
\end{equation}
It can be decomposed as
\begin{equation}
\label{eq_lkh}
    \ln \mathcal{L} = (d \vert h(\boldsymbol{\theta}))  - \frac{1}{2} (h(\boldsymbol{\theta}) \vert h(\boldsymbol{\theta})) -\frac{1}{2} (d \vert d),
\end{equation}
where the last term can be neglected since it does not depend on the estimated parameters and represents a multiplicative constant in the likelihood. The full log-likelihood is a sum over the log-likelihoods of the uncorrelated instrumental channels $A, E, T$. We use an adaptative heterodyned likelihood to speed up the likelihood computation (see~\cite{lazy_lkh, Cornish_2021} and references therein for more information).

Given a chosen prior, the Bayes factor is a means to compute the preference of the data towards one model or another. It is defined as the ratio of the evidence $\mathcal{Z}$ of two models $i$ and $j$
\begin{equation}
    \mathcal{B} = \frac{\mathcal{Z}_i}{\mathcal{Z}_j},
\end{equation}
where the evidence is the integral of the likelihood over the whole parameter's hyper-volume, 
\begin{equation}\label{eq:evidence}
    \mathcal{Z} = \int_{\Theta} \mathcal{L} (\theta) \pi (\theta) d\theta.
\end{equation}

\section{Study of the modes contributions to the SNR}\label{sec4}

The SNR builds up in time and frequency and is defined by $\rho$ as
\begin{equation}
    \rho^2 =  \sum_{lm} \sum_{l'm'} \sum_{I} 4 \Re \int  \frac{\mathcal{H}^{I}_{lm} (f) {\mathcal{H}^{I}_{l'm'}}^*(f)}{S_{n} (f)} df,
\end{equation}
where the sum over independent channels (index $I$) extends over $A,E$ (which have the same noise PSD $S_n$). The assumed PSD is drawn from the `science requirement model' SciRDv1  \cite{SciRD}, including the galactic white dwarf confusion noise, subtracting sources over one year. Note that we are using $\mathcal{H}^{I}_{lm}$ instead of $h_{lm}$, because we also include LISA instrumental response and TDI post-processing combining Eqs.~\eqref{eq_transfer} to \eqref{eq_channels}. They read
\begin{subequations}\label{hlm_tdi}
\begin{align}
    \mathcal{H}^{a}_{lm}(f) = & \, h_{lm}(f)\, \cdot  \frac{i \sqrt{2} \sin(2\pi f L)}{e^{-2 i\pi f L}} \cdot \notag \\ 
    &\left[ (1 + z(f)) (\mathcal{T}_{13}^{lm}(f) + \mathcal{T}_{31}^{lm}(f)) \right. - \notag\\
    & \mathcal{T}_{32}^{lm}(f) - z(f) \mathcal{T}_{23}^{lm}(f) \, - \notag \\
    & \left.\mathcal{T}_{12}^{lm}(f) - z(f) \mathcal{T}_{21}^{lm}(f) \right],\\
    \mathcal{H}^{e}_{lm}(f) = & \,  h_{lm}(f) \, \cdot \frac{i \sqrt{2} \sin(2\pi f L)}{\sqrt{3}e^{-2 i\pi f L}} \cdot \notag \\ 
    &\left[ (1 - z(f)) (\mathcal{T}_{31}^{lm}(f) + \mathcal{T}_{13}^{lm}(f)) \, + \right.\notag \\
    & (2+z(f)) (\mathcal{T}_{21}^{lm}(f) - \mathcal{T}_{23}^{lm}(f)) \, + \notag \\
    & \left. (1+2z(f)) (\mathcal{T}_{12}^{lm}(f) - \mathcal{T}_{32}^{lm}(f)) \right],
\end{align}
\end{subequations}
where $z(f) = e^{2i\pi f  L}$. For convenience, we use the notation of inner product for modes,
\begin{equation}
\label{lmlm}
    (lm\vert l'm') = \sum_{I} 4 \Re \int  \frac{\mathcal{H}^{I}_{lm} (f) {\mathcal{H}^{I}_{l'm'}}^*(f)}{S_{n} (f)} df.
\end{equation}
Then the squared SNR can be written as
\begin{equation}
\label{eq_rho}
    \rho^2 =  \sum_{lm} \sum_{l'm'} (lm\vert l'm').
\end{equation}

\begin{figure}[t]
    \centering
    \includegraphics[width = 0.48\textwidth]{ 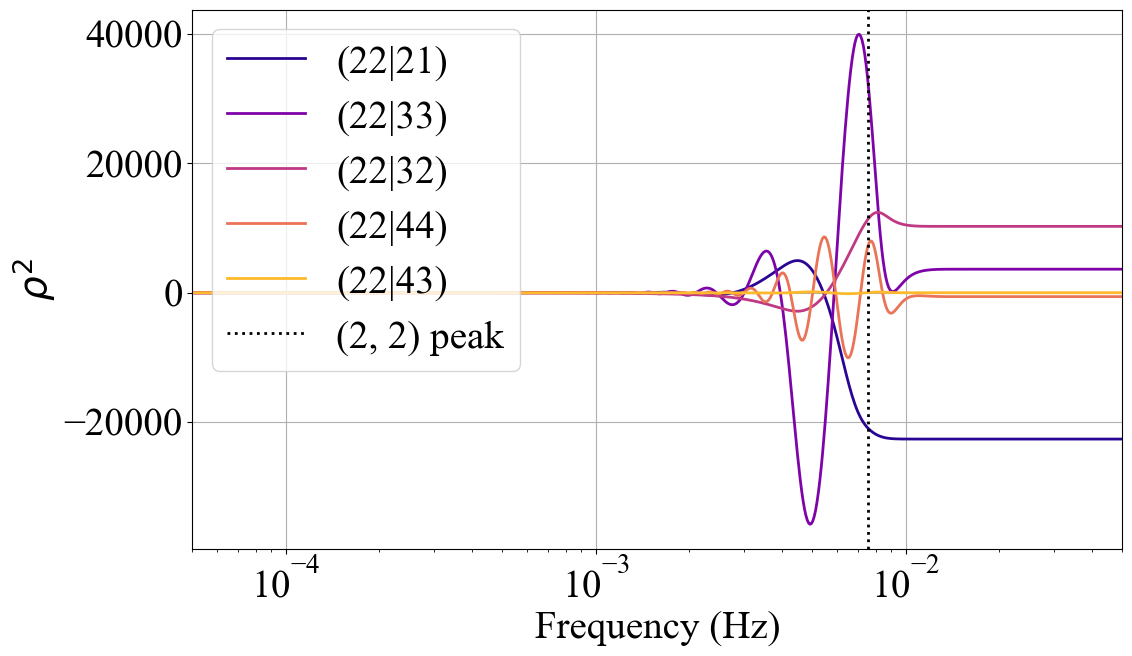}
    \caption{Contribution of cross-terms $(22 \vert l'm')$ to $\rho^2$. We can observe how the cumulative squared SNR changes from positive to negative values and vice-versa until after the ringdown, where it remains constant. The frequency peak of the mode $(2, 2)$ is shown here for guidance. Each mode peaks at a different frequency, so the stabilization period starts at a different point for each pair of modes.}
    \label{fig:snr_freq_22}
\end{figure}

In Eq.~\eqref{lmlm}, the cross-terms $(lm\vert l'm')$ have no reason to be positive and can contribute negatively to the total SNR. In other words, the phases can be constructive or destructive, which depends on the values of the ecliptic latitude $\beta$ and longitude $\lambda$, the inclination $\iota$, the phase $\phi$, the polarization angle $\psi$ and the mass ratio $q$. An illustration of this outcome can be seen in Fig.~\ref{fig:snr_freq_22}. For this example, we use the parameters written in Table~\ref{tab:sources} with a redshifted total mass of $\sim 2.44\times 10^6$ M$_\odot$. 
In Fig.~\ref{fig:snr_freq_22}, we show the $(22\vert l'm')$ cross-terms of the accumulated squared SNR $\rho^2$ varying between positive and negative values depending on the frequency. After a certain point, it remains constant since there is no more contribution to the SNR, neither positive nor negative. This happens at a different frequency for each pair of modes. We plot the frequency peak of the $(2, 2)$ mode with a dotted black line as a guide.

\begin{ruledtabular}
\begin{table}
    \centering
    \caption{Source parameters in SSB frame with aligned spins and redshifted masses.} 
    \begin{tabular}{cc|cc}
    Parameter & Value & Parameter & Value\\
    \hline
        Mass
        (M$_{\odot}$) & $ [10^5, 5\times10^9]$ & $\beta$ (rad) &  $ \pi/2 $ \\
        q (Mass ratio) &  $2$ & $\lambda$ (rad) & $\pi$ \\
        $\chi_1$ & $0.5 $ & $\phi$ (rad) & $\pi/2$ \\
        $\chi_2$ & $0.5 $ & $\psi$  (rad)& $\pi/2$ \\
        redshift  & 3 & $\iota$  (rad) & $\pi/3$ \\
    \end{tabular}
    \label{tab:sources}
\end{table}
\end{ruledtabular}

\begin{figure*}
    \centering
    
    \includegraphics[width = 0.84\textwidth]{ 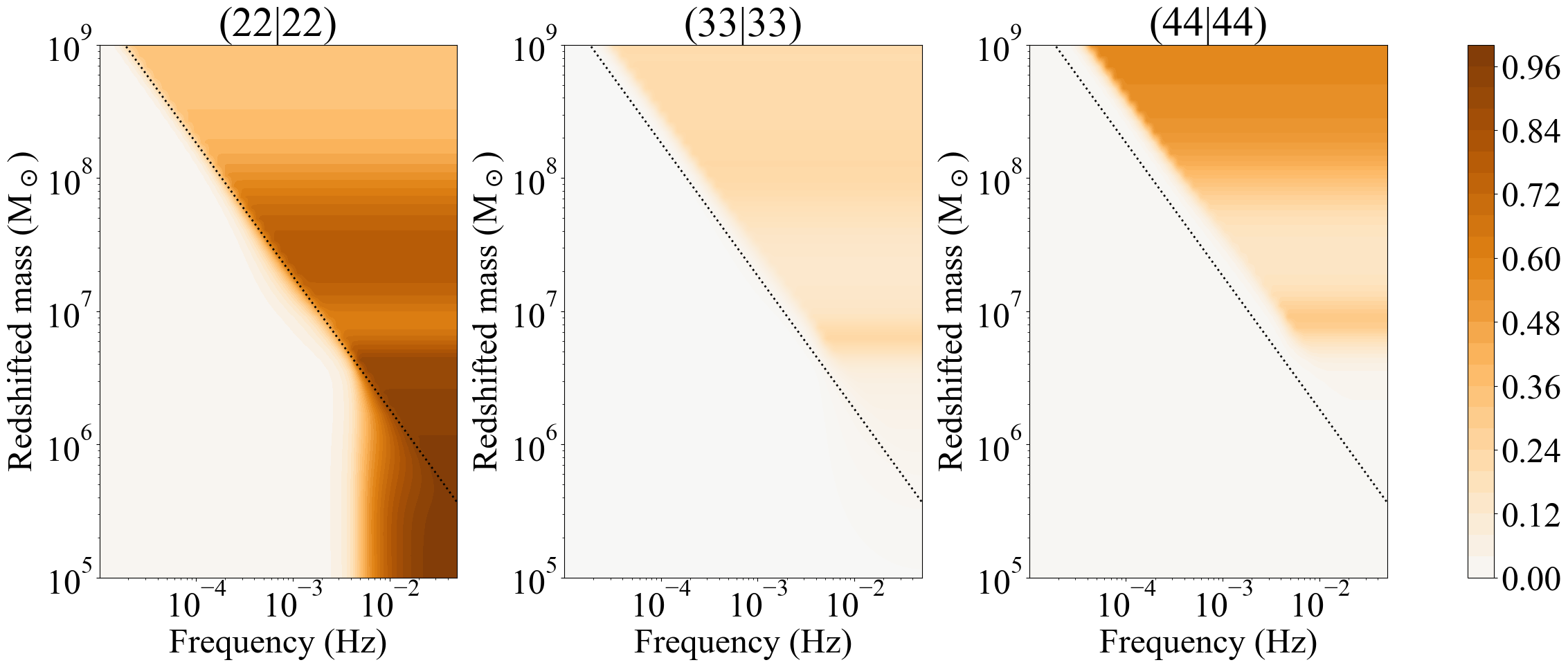} \hypertarget{fig:snr_diag_sym}{}
    \caption*{ a) Ratio contribution of the square terms $(lm \vert lm)$ to the total squared SNR.}
    
    \includegraphics[width =0.86\textwidth]{ 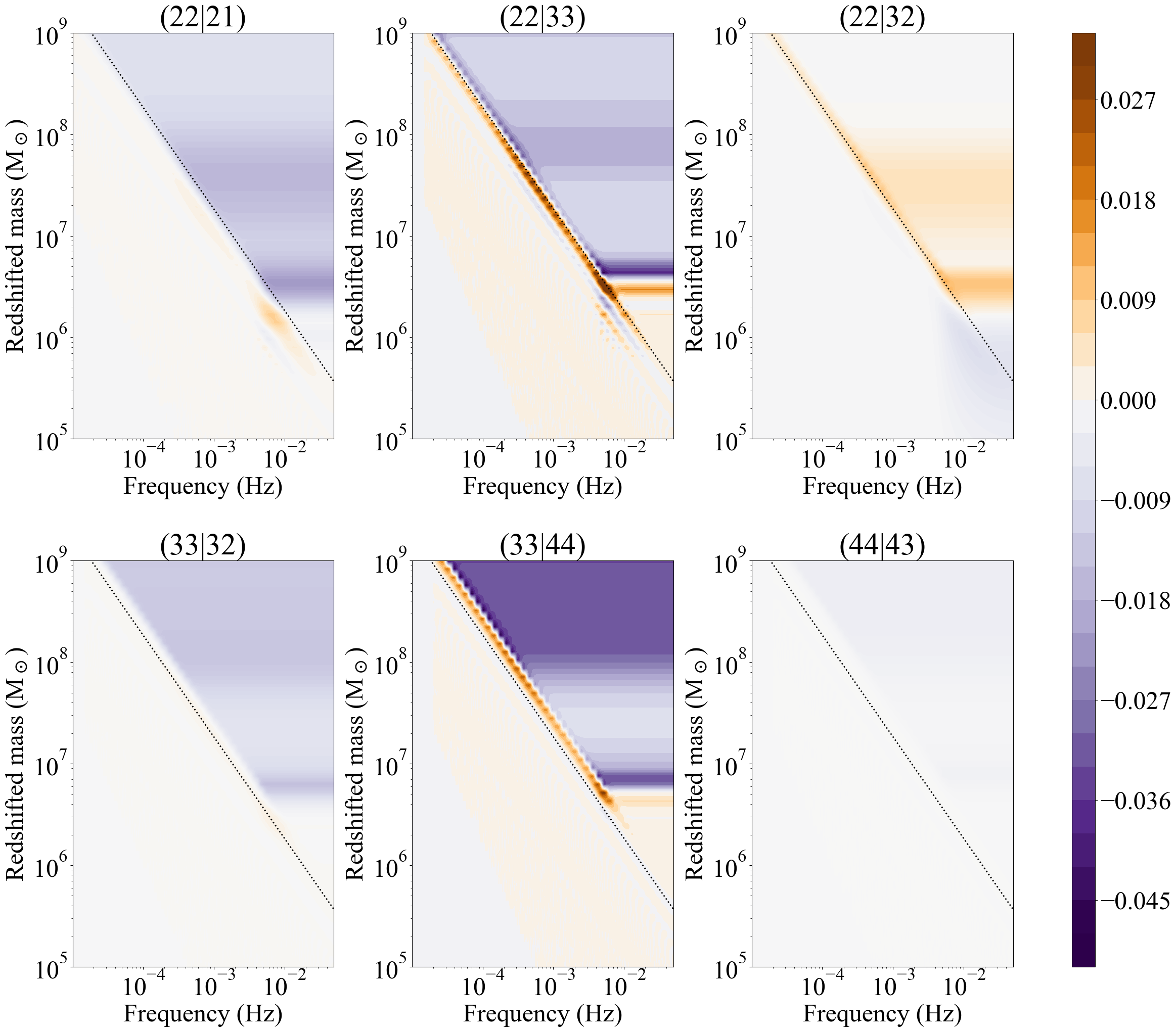} \hypertarget{fig:snr_diag_most}{}
    \caption*{ b) Highest ratio contributions of the cross-terms $(lm \vert l'm')$ to the total squared SNR.}
    
    \caption{Cumulative contribution to the squared SNR of pairs of modes depending on the total redshifted mass. In the top row, we find the square terms with the largest ratios. The quadrupolar square term represents most of the contribution for low-mass sources, while it slowly decreases for more massive ones. The opposite happens for terms $(33 \vert 33)$ and $(44 \vert 44)$ albeit to a smaller extent. We show the cross-terms in the last two rows. They oscillate between $\sim 1 \%$ and $5 \%$, especially near the ringdown, illustrated with a dotted line for the mode $(2, 2)$ as guidance. This percentage can represent an SNR of tens for a total SNR of thousands.}
    \label{fig:snr}
\end{figure*}

\subsection{Dependency on mass and frequency}

The next natural step to understanding the detectability of modes is to analyze the contribution of each pair of modes to the total SNR for different sources. In general, the SNR depends on all parameters of the source. To simplify our analysis, we fix most parameters to arbitrary values (listed in Table~\ref{tab:sources}) and let the mass vary. Note that the results, particularly the details of the hierarchy between modes and cross-terms, might depend on this choice of parameters. For each source's total mass and each frequency bin, we compute the accumulated SNR for each mode and normalize it to the total SNR. Since the ratio of the total mass over the luminosity distance increases for higher masses, the normalization allows us to compare the contribution of each mode regardless of their absolute SNR value.

We gather the results in a set of plots in Figs.~\hyperlink{fig:snr_diag_sym}{\ref{fig:snr}a} and~\hyperlink{fig:snr_diag_most}{\ref{fig:snr}b}. Each panel of the figure corresponds to a mode and shows the squared SNR dependence on both the source mass and the observed frequency. Then, for a given mass, we can observe how the squared SNR accumulates in frequency for each pair of modes. Each set of plots has its color bar where warmer colors correspond to larger accumulated SNR contributions. The numbers at the right of the bars represent the contribution ratio of each pair of modes to the total squared SNR $\rho^2$. Note that we use squared SNR instead of SNR so that the sum of all contributions is equal to 1. This choice also allows us to highlight the negative contributions we mentioned before and their direct impact on the likelihood (see Eqs. \eqref{eq_lkh} and \eqref{eq_rho}). That said, the most noticeable feature is the positive contribution of the square terms $(lm \vert lm)$ over all the frequencies, while in contrast, the cross-terms $(lm \vert l'm')$ can have negative contributions. We decided to plot the frequency peak of the $(2, 2)$ mode (diagonal dotted line) as a mapping guide since the contribution to the SNR changes considerably in the merger-ringdown regime. Depending on the mode, the SNR variation will start before or after this frequency line. 

In the left plot of Fig.~\hyperlink{fig:snr_diag_sym}{\ref{fig:snr}a}, we note that the contribution of $(22 \vert 22)$ to $\rho^2$ is between $80 \%$ to  $94 \%$ of the total, up to masses $\sim 4\times 10^6$ M$_{\odot}$, while it decreases to $30 \%$ around masses of $\sim 10^8$ M$_{\odot}$. The rectangular darker area at the bottom right, between frequencies [$10^{-3} - 5\times 10^{-2}$] Hz and masses [$10^5 - 2\times 10^6$]~M$_\odot$, indicates that most of the SNR comes from the inspiral part. This is expected since the waveform peaks outside or at the limit of the LISA frequency band. The pair $(33 \vert 33)$ (in the center) has a small contribution for low masses but exhibits a considerable increase up to $21 \%$ for high masses around $10^8$ M$_{\odot}$. Finally, the case of $(44 \vert 44)$ (right plot) shows a similar behavior but is augmented by a factor $\sim 2.5$, representing about $54 \%$ of the total SNR for large-mass SMBHBs. In other words, if the redshifted total mass of the system is larger than $10^8$ M$_{\odot}$, the contribution of the quadrupolar mode will no longer predominate. It can represent half of the mode $(4, 4)$ and only a factor of 1.5 bigger than the mode $(3, 3)$. This highlights the importance of including higher harmonics to describe SMBHB signals. 

The second set of plots in Fig.~\hyperlink{fig:snr_diag_most}{{\ref{fig:snr}b}} shows the pairs of modes that make the highest contributions to the squared SNR, representing approximately from $1 \%$ to $5 \%$ of the total. This percentage might seem small, but for SMBHBs with a total SNR of 1000, the cross-term contribution can be from 10 to 50. Those contributions are even higher than other square terms not shown here. Even though the contributions can be negative, we use their absolute value for comparison. The reason is that the sign depends on the extrinsic parameters and the mass ratio, which vary from source to source, but are fixed here to arbitrary values for the sake of an illustration.

The position of the mode's frequency peak relative to the LISA sensitivity curve is driven by the event mass. Therefore, the SNR of higher modes can become more relevant than the $(2,2)$ mode for large mass events.

\subsection{Dependency on source mass and redshift}\label{sec_mass}

\begin{figure*}
    \centering
    \includegraphics[width =0.7\textwidth]{ 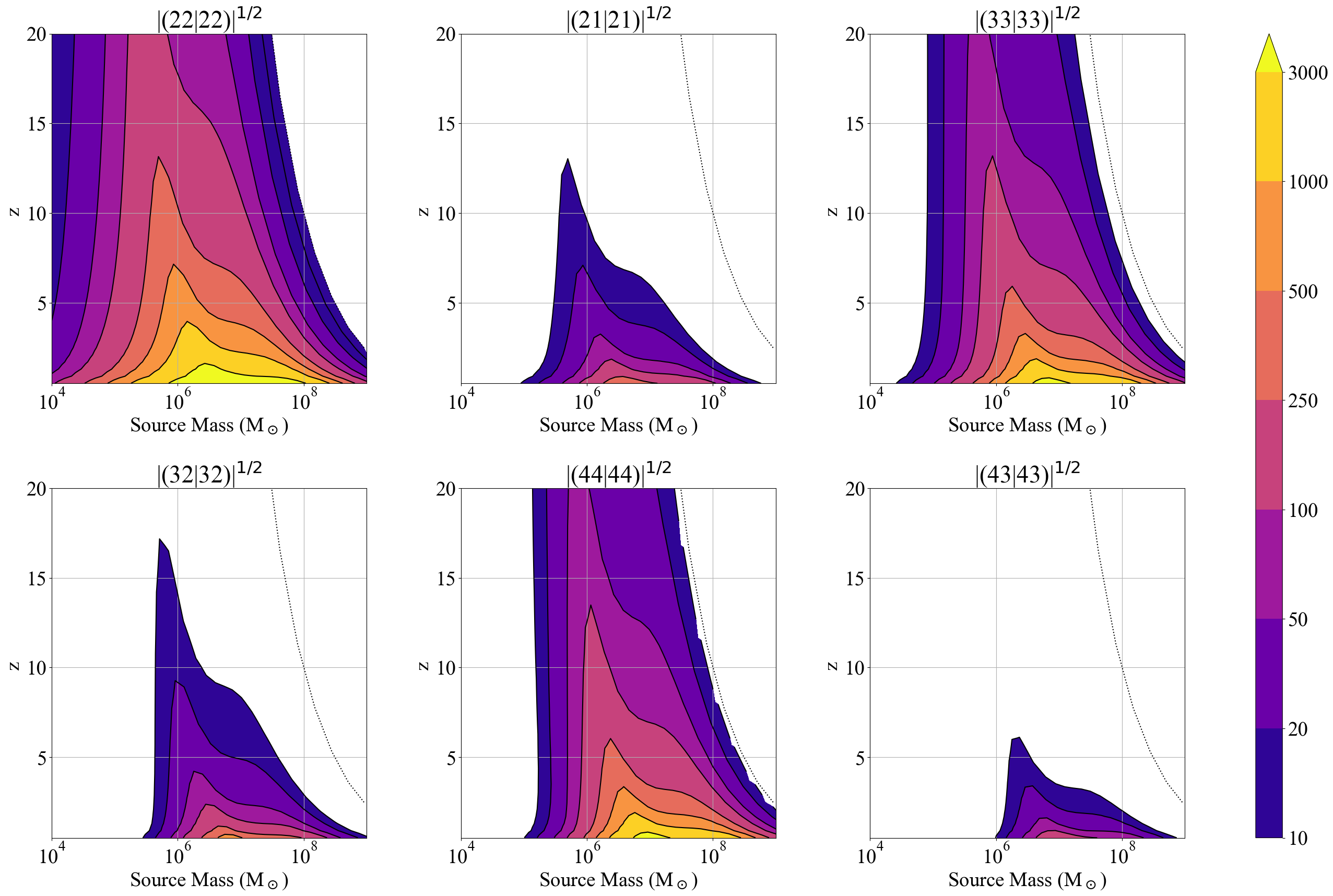}\hypertarget{fig:waterfall_sym}{}
    \caption*{a) Contour plot with the contributions of square pairs of modes $\vert (lm \vert lm)\vert ^{1/2}$.}
    
    \includegraphics[width= 0.7\textwidth]{ 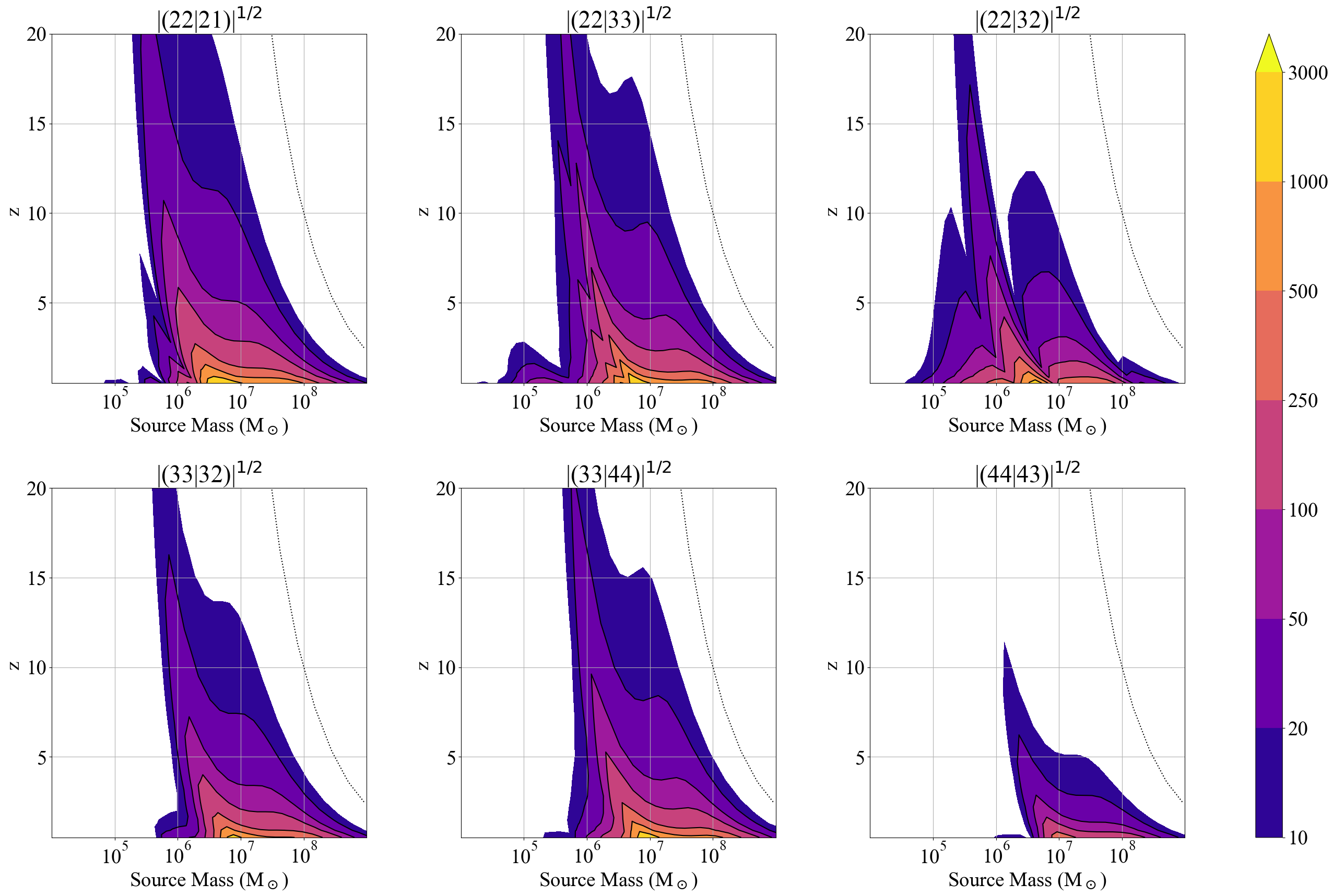}\hypertarget{fig:waterfall_most}{}
    \caption*{b) Contour plot with the highest SNR contributions of cross-terms $\vert (lm \vert l'm')\vert ^{1/2}$.}
    
    \caption{Contour plot for the root squared absolute cumulative value of contributions $\vert ( lm \vert l'm' ) \vert ^{1/2}$. The dotted line corresponds to the total SNR with a $\geq 10$ threshold, plotted here for comparison. We remark the sub-dominant contribution of  $\vert(33 \vert 33)\vert ^{1/2}$ and $\vert(44 \vert 44)\vert ^{1/2}$ after the quadrupolar square term. Note the 0-contributions in some cross-terms due to representing a single system with fixed parameters. The complete array for all contributions is plotted in Fig.~\ref{fig:waterfall_all} of the Appendix~\ref{app_1}.} 
    \label{fig:waterfall}
\end{figure*}

To represent the impact of the mass and the distance on the SNR, as mentioned in Section~\ref{sec3}, we show in Figs.~\hyperlink{fig:waterfall_sym}{\ref{fig:waterfall}a} and \hyperlink{fig:waterfall_most}{\ref{fig:waterfall}b} the contour plots for different modes depending on the total mass in the source frame. We choose a source with parameters listed in Table~\ref{tab:sources} for the sake of an illustration. However, the contributions would change with different parameters. Note that the lower bound of the total SNR is 10, which is the adopted threshold for SMBHB detection in LISA~\cite{SciRD, LISA_Proposal2017}.

In these figures, we can see the high contribution of the square terms $\vert(l, m=l \vert l,m=l) \vert ^{1/2}$ as well as the highest-contributing cross-terms such as $\vert (22 \vert 21)\vert ^{1/2}$, $\vert(22 \vert 32)\vert ^{1/2}$ and $\vert(33 \vert 32)\vert ^{1/2}$. The complete array of pairs of modes is shown in Fig.~\ref{fig:waterfall_all} of Appendix~\ref{secA2}. The pairs $\vert (22 \vert 33)\vert ^{1/2}$ and $\vert(22 \vert 32)\vert ^{1/2}$ exhibit a drop in SNR for systems with a total source mass around $10^6$ M$_\odot$, depending on the redshift. This effect results from the specific choice of parameters of the source we study. It is also visible in Fig.~\ref{fig:snr_freq_22}, where some cross-terms have a null cumulative contribution to the total squared SNR.

We also investigated the effect of the mass ratio on mode SNR by performing the same analysis, observing the SNR cancellation expected for equal-mass systems (see Appendix \ref{secA2}). Contour plots are shown in Fig.~\ref{fig:waterfall_q_sym} also in Appendix~\ref{secA2}.

\section{Example of Bayesian analysis}\label{sec5}
\subsection{Data and models}\label{models}

To understand the impact of higher modes in the parameter inference, we inject a SMBHB source signal with 6 modes. The parameters used to generate this source were taken from the LISA Data Challenges (LDC) \textit{Sangria}'s catalog \cite{le_jeune_maude}, and detailed in Table~\ref{tab:params}, where the subscript $L$ in the extrinsic parameters means that it is expressed in the LISA frame at the time of coalescence $t_c$. We also use the redshifted chirp mass $M_c$ and mass ratio q~$\geq~1$ instead of the individual masses. The parameters $\chi_i$ represent the aligned (non-precessing binaries) BH's dimensionless spin and $D_L$ is the luminosity distance.

\begin{ruledtabular}
\begin{table}
    \centering
    \caption{Parameters of the SMBHB source with SNR 744 chosen from \textit{Sangria}'s LDC catalog for redshifted mass without precession, along with the flat priors intervals used in the inference.}
    \begin{tabular}{ccc}
    Parameter & Value & Prior\\
    \hline
    M$_c$ ($M_{\odot}$) & 857080.8396 & $ [10^4, 5\times10^7]$ \\
    q & 2.7589  &  $[1, 10]$ \\
    $\chi_1$ & -0.5488 & $[-1,\,  1]$ \\
    $\chi_2$ & 0.2317 & $[-1,\,  1]$ \\
    $D_L$ (Mpc) & 40084.6792 & $[10^4, \, 5\times10^6]$ \\
    $t_c$ (s) & 0.0 & $[- 600,\,  600]$ \\
    $\beta_L$ (rad) & -0.6186 & $[-\pi/2, \, \pi/2]$ \\
    $\lambda_L$ (rad) & 2.2782 & $[0, \, 2\pi]$ \\
    $\phi$  (rad)& 0.2492 & $[-\pi, \pi]$ \\
    $\Psi_L$ (rad) & 1.5158 & $[0, \pi]$ \\
     $\iota$  (rad) & 2.5969 & $[0, \pi]$
    \end{tabular}
    \label{tab:params}
\end{table}
\end{ruledtabular}

Two data sets were considered, one without noise and another with noise. The noise was generated with the PSD assumed in Section~\ref{sec4}. Each data set includes two TDI channels, $A$ and $E$. In both cases, we restricted the frequency band to the interval $[10^{-5} - 5\times 10^{-2}]$ Hz and ran a nested sampling algorithm to estimate the source parameters for various models. We chose the sampler \texttt{dynesty}~\cite{dynesty}, as it allows us to obtain approximate evidence estimates (see Section~\ref{sec_Bayes} and particularly Eq.~\ref{eq:evidence}). As a consistency check, we also ran \texttt{ptemcee}, a parallel-tempered Markov Chain Monte Carlo ensemble sampler~\cite{Vousden_2015,Foreman_Mackey_2013}. We obtained consistent results with the two samplers, with a slightly better convergence for \texttt{ptemcee} (which does not allow for direct evidence computation, however). We report \texttt{dynesty} results in the following.  

We consider 6 models for the parameter estimation, where each one describes the waveform with a certain number of modes. The first model generates the gravitational signature with only the quadrupolar mode $(2, 2)$. The other models (see Table~\ref{tab:models} for models' definition) include an increasing number of higher harmonics, ranked by their SNR contribution $(lm \vert lm)$, as observed in Fig.~\hyperlink{fig:waterfall_sym}{\ref{fig:waterfall}a}. This amounts to first selecting successive $(l, m=l)$ modes with increasing $l$ and then the $(l, m=l-1)$ modes.

\begin{ruledtabular}
\begin{table}
    \centering
    \caption{Each model is indexed according to the number of modes included in the waveform generation.}
    \begin{tabular}{cc}
        Model & Modes (l,m) \\
        \hline
        $M_1$ & (2, 2)\\
        $M_2$ & (2, 2), (3, 3)\\
        $M_3$ & (2, 2), (3, 3), (4, 4)\\
        $M_4$ & (2, 2), (3, 3), (4, 4), (2, 1)\\
        $M_5$ & (2, 2), (3, 3), (4, 4), (2, 1), (3, 2)\\
        $M_6$ & (2, 2), (3, 3), (4, 4), (2, 1), (3, 2), (4, 3)\\
    \end{tabular}
    \label{tab:models}
\end{table}
\end{ruledtabular}

The priors we use are flat for all parameters in the intervals written in Table~\ref{tab:params}, except for the chirp mass, which has a uniform prior in logarithmic scale. We use the whole physically allowed interval for the extrinsic parameters, while we use a raw estimation of the expected values for the intrinsic ones. We use a narrow prior for the coalescence time as it can easily be spotted in the detection process but with a difference of up to 600 seconds between the LISA and the SSB reference frame. Note that the polarization $\psi$ is allowed to go from 0 to $\pi$ (and not $2\pi$) to prevent parameter degeneracy, given that in the antenna pattern, the polarization is always preceded by a factor 2 as shown in Eq.\eqref{eq:pol}.


\subsection{SNR and Bayes factor}\label{sec_snr}

Before presenting the Bayes factor and parameter estimation results, we discuss our expectations regarding the contribution of modes. Our example considers a source event randomly chosen from the LDC Sangria's catalog with a SNR $\sim 744$ (see Table~\ref{tab:params}). Converting the redshifted total mass to the source-frame total mass with a redshift of 4.3, we obtain the value of $2.28 \times 10^5$ M$_\odot$. Using Fig.~\ref{fig:waterfall}, with this source-frame mass, we can expect the term $(22\vert 22)$ to be the dominant contributor, whereas the rest of the SNR will come from $(33\vert 33)$ and $(44\vert 44)$ and the cross-terms $(22\vert 21)$ and $(22\vert 32)$. This is a simple estimation since the source parameters are not precisely the same as the ones listed in Table~\ref{tab:sources}.

\begin{ruledtabular}
\begin{table}
    \centering
    \caption{Bayes factor for all models compared to the injected model}
    \begin{tabular}{ccc}
    Bayes factor & Noiseless data set  & Noisy data set\\
    \hline
        log($\mathcal{Z}_1/\mathcal{Z}_6$) & -6845  & -6873\\
        log($\mathcal{Z}_2/\mathcal{Z}_6$) & -976  & -1015\\
        log($\mathcal{Z}_3/\mathcal{Z}_6$)& -237  & -259\\
        log($\mathcal{Z}_4/\mathcal{Z}_6$) & -109  & -134\\
        log($\mathcal{Z}_5/\mathcal{Z}_6$) & -84  & -100\\
    \end{tabular}
    \label{tab:bayes}
\end{table}
\end{ruledtabular}

\begin{figure}[t]
    \centering
    \includegraphics[width=0.45\textwidth]{ 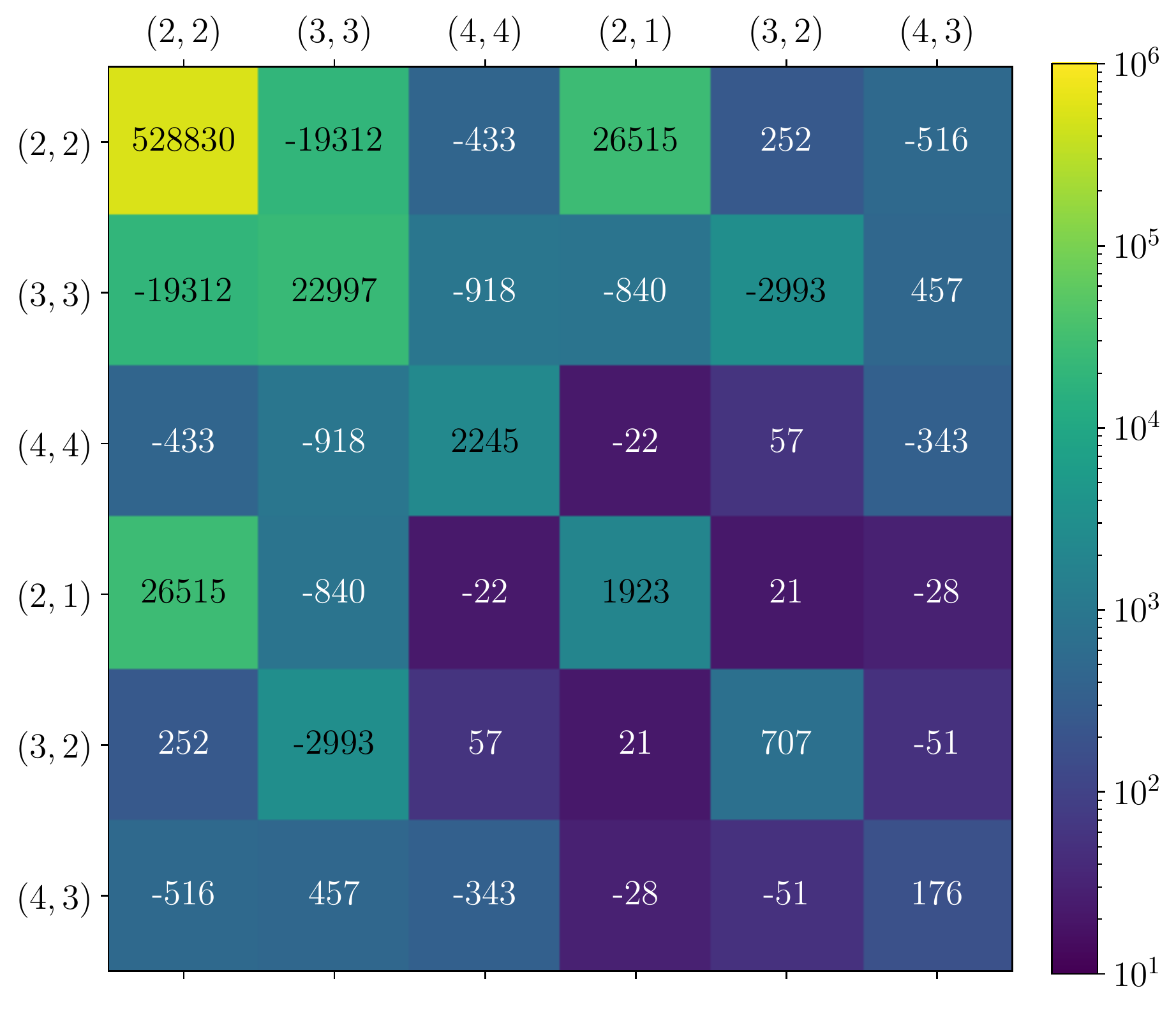}
    \caption{Final value of squared SNR of each pair of modes $(lm \vert l'm')$ for our example. Note the positive values for the square terms and the decreasing values in the diagonal. In contra-position, note the negative values for pairs with different `m' except for cross-terms $(22 \vert 21)$ and $(44 \vert 32)$. An interesting result is the relatively high value of $(22 \vert 21)$, the second highest value.}
    \label{fig:snr_matrix}
\end{figure}

\begin{figure*}
\centering
\begin{tabular}{cc}
    \includegraphics[width = 0.45\textwidth]{ 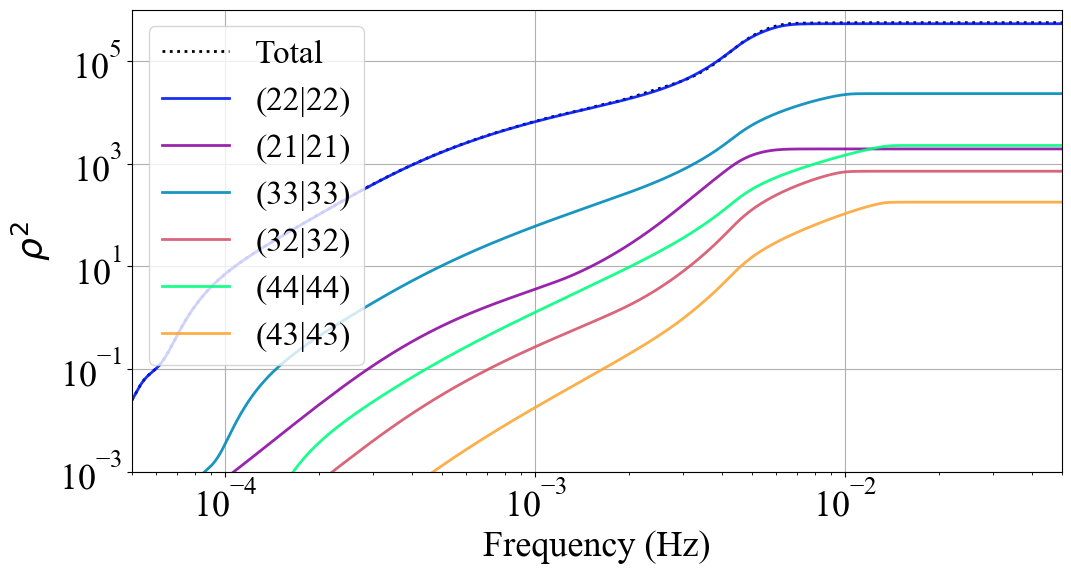}
    &
    \includegraphics[width =0.45\textwidth]{ 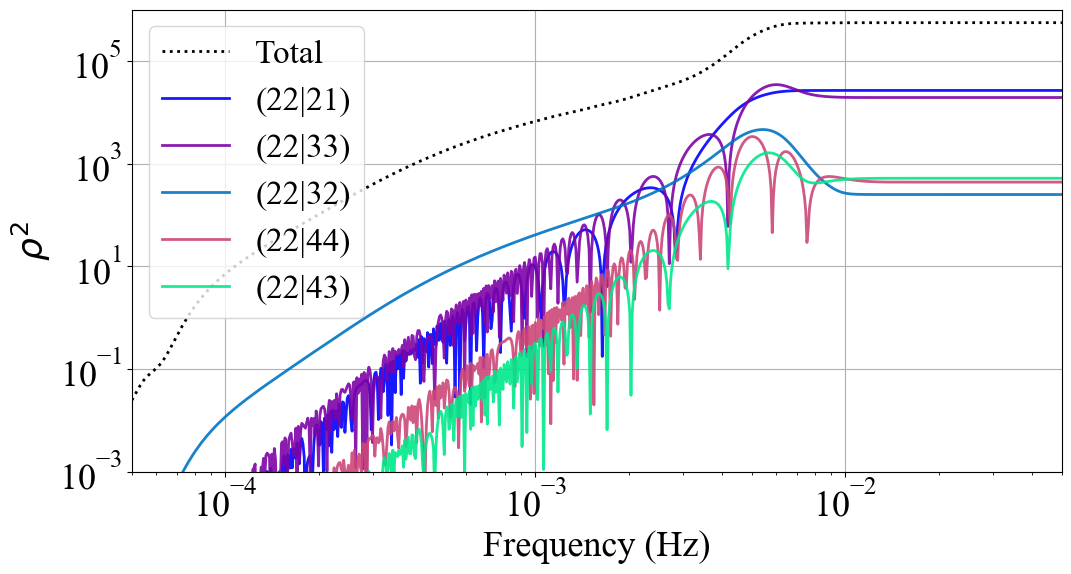}
    \\
    a) square terms $(lm \vert lm)$ &
    b) Cross-terms $(22 \vert l'm')$ \\
    \\
    \includegraphics[width = 0.45\textwidth]{ 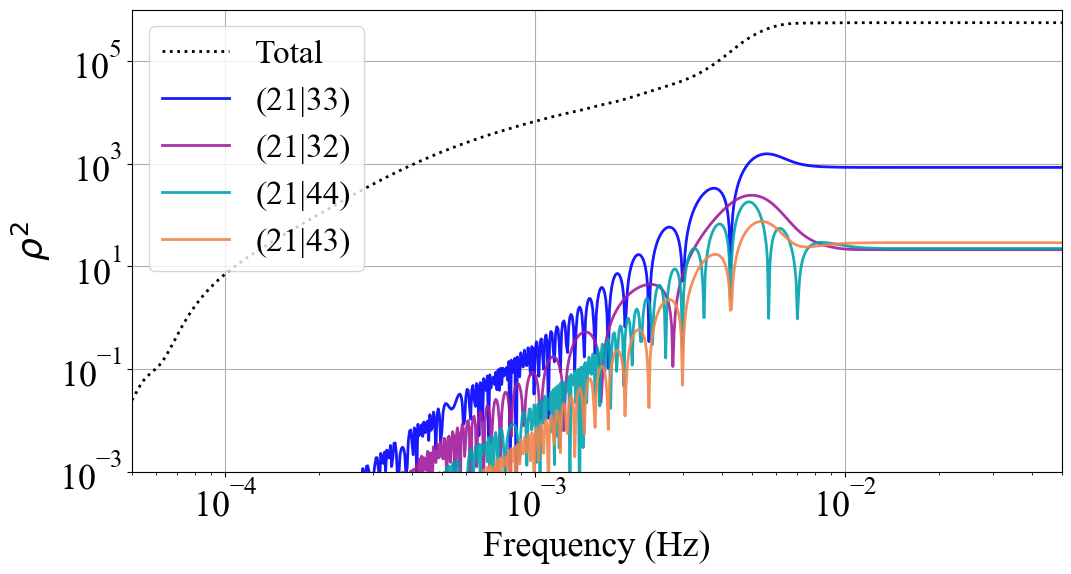}
    &
    \includegraphics[width = 0.45\textwidth]{ 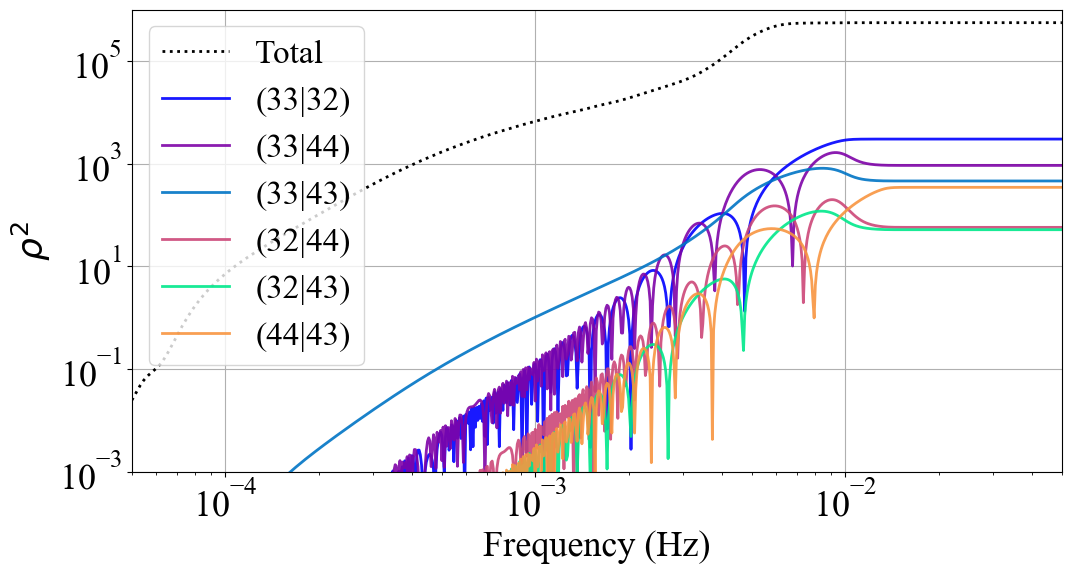}
    \\
    c) Cross-terms $(21 \vert l'm')$ &
    d) Cross-terms $(lm \vert l'm')$
    \end{tabular}
    \caption{Cumulative contribution to the squared SNR (in absolute value) of different pairs of modes depending on the frequency. The upper left figure shows the contribution of the square pairs, where we can see the quadrupolar making the higher contribution. All the other figures show cross-terms, from where we can highlight the contribution of the quadrupolar mode with higher modes $(22 \vert l'm')$, which are predominant over other cross-terms.}
    \label{fig:snr2}
\end{figure*}

The actual contribution from each pair of modes is plotted in Fig.~\ref{fig:snr_matrix}, where we show the squared SNR contribution for each pair of modes for the full IMR signal of our example source. Both axes correspond to the modes, so the intersection represents the pair of modes $(lm\vert l'm')$. In the diagonal of the matrix, we find the square terms, while in the upper and lower triangle, we encounter the symmetric cross-terms. The value in each box indicates the squared SNR of each pair, whose absolute value is shown by the color bar. We can observe that the pair $(22\vert 22)$ indeed accounts for the largest contribution as expected, followed by the pairs $(22 \vert 21)$, $(33 \vert 33)$, and $(22 \vert 33)$. The color gradient we observe when descending in the diagonal line is a consequence of the hierarchic ordering of modes. One striking difference with Fig.\ref{fig:waterfall} is the high contribution of $(22 \vert 21)$ when compared to $(33 \vert 33)$, showing that the details of the mode contributions will vary for different sources with different intrinsic and extrinsic parameters.

Even if the final value of a mode's SNR is small for the complete IMR signal, it does not mean that its impact is negligible in relative terms everywhere in frequency, particularly in the pre-merger phase. This feature can be observed in Fig.~\ref{fig:snr2}, where we represent the squared SNR absolute value as a function of the frequency. The pairs' contributions are separated into groups to make the figures readable. We can see how some terms dominate in their group below a frequency that approximately corresponds to the merger, after which they later decrease. This happens for the pair $(22 \vert 32)$ in the top-right figure or $(21 \vert 32)$ in the bottom-left figure. Thus, statements about modes' relative importance generally depend on the total accumulated SNR.

To quantify our ability to identify the presence of modes, we compute the Bayes factor using the \texttt{dynesty} sampler. We compare all the models $M_k$ with $k=1,\hdots,5$ with $M_6$. The results gathered in Table~\ref{tab:bayes} show clear negative values for all of them. This means the model with 6 modes is preferred and describes the data better than all other models, as expected. Even the value of -84 (-100 with noise) shows a significant preference for the model $M_6$ over $M_5$, where only the mode (4, 3) is absent. Thus, even the weakest modes in our setting should be identified as present in the data, which indicates that LISA observations will be capable of identifying waveform modes beyond the ones available in current waveform models. This result advocates using waveforms with all available higher harmonics to capture all the physics in LISA signals and further developing waveforms with higher mode contents. In the following section, we investigate whether ignoring these weaker higher modes would produce biased parameter estimation results.

\begin{figure*}
    \centering
    \includegraphics[width=0.55\textwidth]{ 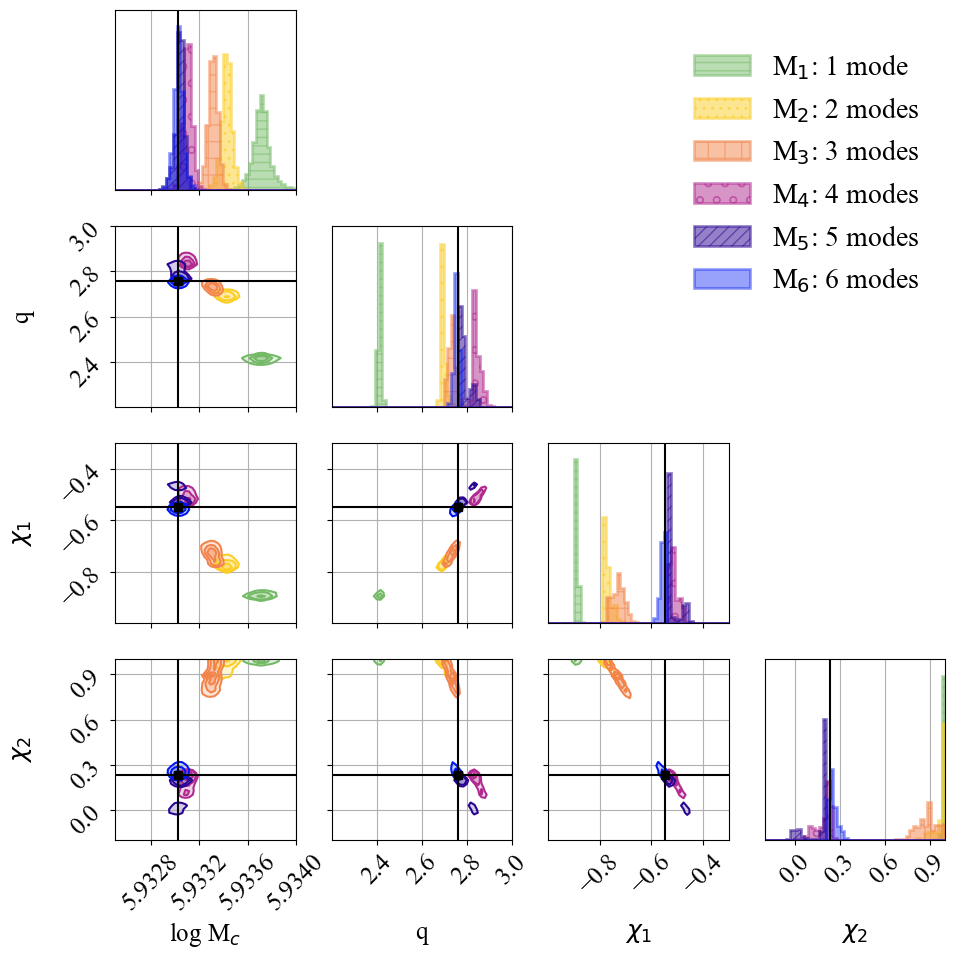}
    \caption*{a) Marginalized posterior distribution on mass and spin parameters without noise. }
    \includegraphics[width=0.55\textwidth]{ 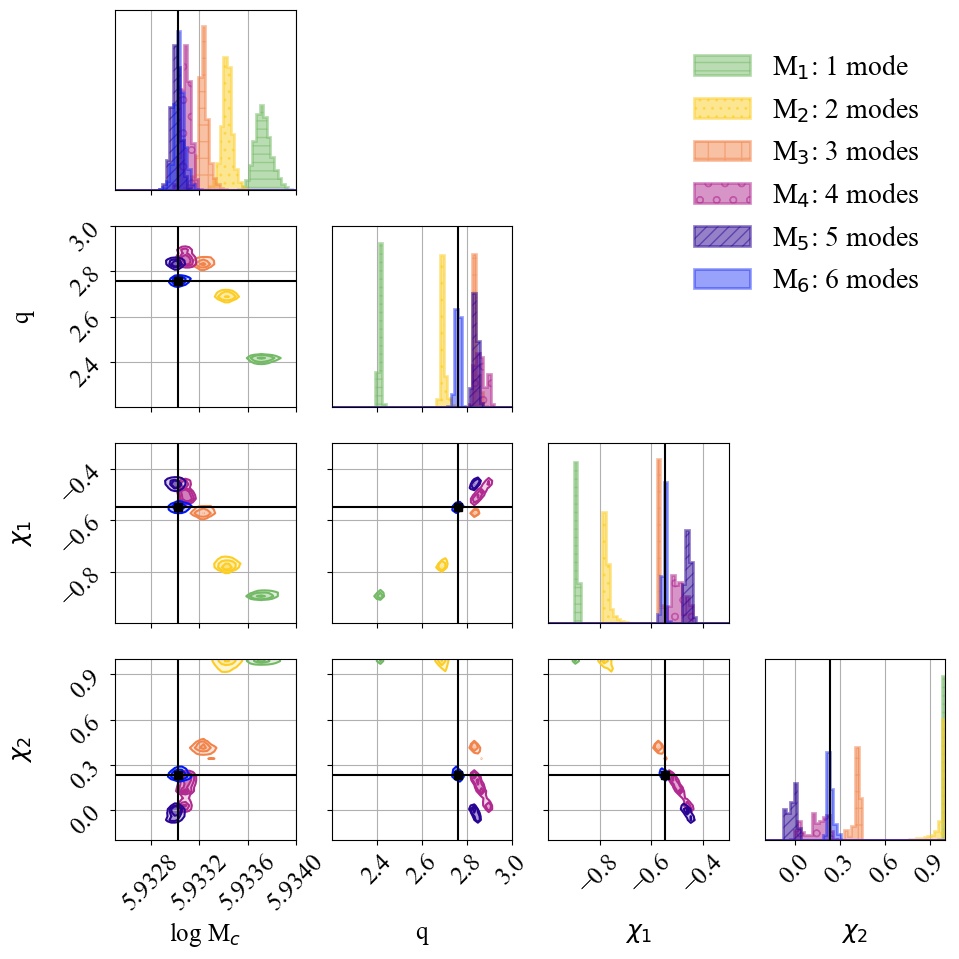}
    \caption*{b) Marginalized posterior distribution on mass and spin parameters for noisy data.}
\caption{The six models with different numbers of modes are represented here. The posterior of the model $M_6$ (blue) finds the true values with high accuracy, while other models tend to induce biases, especially for spin parameters. In the bottom figure, we see how the introduction of noise does not decrease our ability to find the true values with 6 modes for this particular source.}
\label{fig:res}
\end{figure*}

\begin{ruledtabular}
\begin{table*}
    \centering
    \caption{Estimated value for models $M_1$ and $M_6$, for the two data sets, without and with noise.}
    \begin{tabular}{cc|cc|cc}
    \multirow{2}{*}{Parameter} & \multirow{2}{*}{True value} & Estimated value & Estimated value  & Estimated value with & Estimated value with \\
    & & with $M_1$ (noiseless) & with $M_6$ (noiseless)&  $M_1$ (with noise) & $M_6$ (with noise) \\
    \hline
    $\log M_c$  ($M_\odot$) & 5.93302 & 5.93371$^{+0.00019}_{-0.00016}$ & 5.93303$^{+0.00010}_{-0.00010}$ & 5.93374$^{+0.00019}_{-0.00016}$ & 5.93304$^{+0.00009}_{-0.00010}$\\
    q & 2.759 & 2.411$^{+0.012}_{-0.011}$ & 2.751$^{+0.019}_{-0.021}$ & 2.414$^{+0.012}_{-0.012}$ & 2.759$^{+0.013}_{-0.023}$\\
    $\chi_1$ &  -0.549 & -0.890$^{+0.009}_{-0007.}$ & -0.559$^{+0.016}_{-0.024}$ & -0.888$^{+0.009}_{-0.008}$ &  -0.549$^{+0.011}_{-0.021}$\\
    $\chi_2$ & 0.232 & 0.996$^{+0.004}_{-0.019}$ & 0.261$^{+0.064}_{-0.043}$ & 0.996$^{+0.004}_{-0.018}$  & 0.231$^{+0.057}_{-0.030}$\\ 
    \end{tabular}
\label{tab:estimated_params}
\end{table*}   
\end{ruledtabular}
\subsection{Posterior and parameter bias}\label{sec_results}

To assess the impact of the mode's contribution on the estimation of the parameters, we show in Fig.~\ref{fig:res} the posterior distribution of the logarithm of the chirp mass $\log M_c$, the mass ratio $q$ and individual adimensional spins $\chi_1$, $\chi_2$. Note that these parameters are relevant to the description of the remnant BH (see Section~\ref{sec1} and \cite{Pan_2011, PhenomD}). The complete array of the parameter's posterior distribution can be found in Fig~\ref{fig:res_noise_all} in Appendix~\ref{secA2}.

We show here the six models for comparison. Intersected black lines represent the true values, and each model's posterior distribution is distinguished by the color code indicated in the legend. The parameter estimation for the model with only the quadrupolar mode (green color) leads to biased estimates. As we increase the number of harmonics in the models, the parameter posterior means get closer to the injected values. We observe that the posterior of model $M_6$ (in blue) is centered on the true value for all parameters, which is expected since the signal is injected and recovered with the same model. By comparing models with 3 and 4 modes ($M_3$ and $M_4$, orange and pink, respectively), we observe a better estimation of spin parameters when the mode $(2, 1)$ is included in the waveform (model $M_4$). This observation is consistent with the large relative contribution of $(22 \vert 21)$ indicated by Fig.~\ref{fig:snr_matrix}. The explanation of the importance of the $(2, 1)$ mode and whether this is generic or specific to our example source are left for future investigations. Note that PhenomHM generates the inspiral phase of the waveform with post-Newtonian approximation, however, the SNR is dominated by the late-inspiral-merger regime. 

In Table~\ref{tab:estimated_params}, we list the parameter's injected values and the estimated values, with models $M_1$ and $M_6$ in the presence and absence of noise. The super- and sub-scripts indicate the $68\%$ confidence interval. In both cases, i.e., with and without noise, the model featuring only the dominant quadrupole mode is inaccurate in finding the true values. In contrast, the estimated mean value with all modes is consistent with the injection. Surprisingly, the values obtained with the noisy data and the $M_6$ template appear closer to the injection than the ones obtained without noise. Note that the posterior distributions are not perfectly Gaussian and the mean value can be shifted due to tails. We performed another run with a different realization of the noisy data, obtaining similar results as the data without noise. The evidences we obtained with both noisy realization encompass the evidence obtained the data without noise. Therefore, we interpret this particular result as a fortuitous outcome of statistical fluctuations. In conclusion, for the medium-SNR and medium-mass case that we study (see Fig.\ref{fig:snr_matrix}), the absence of higher modes would already result in biased estimated values. These biases would only get worse for higher-SNR and higher-mass systems.

\begin{figure}[t]
    \centering
    \includegraphics[width=0.46\textwidth]{ 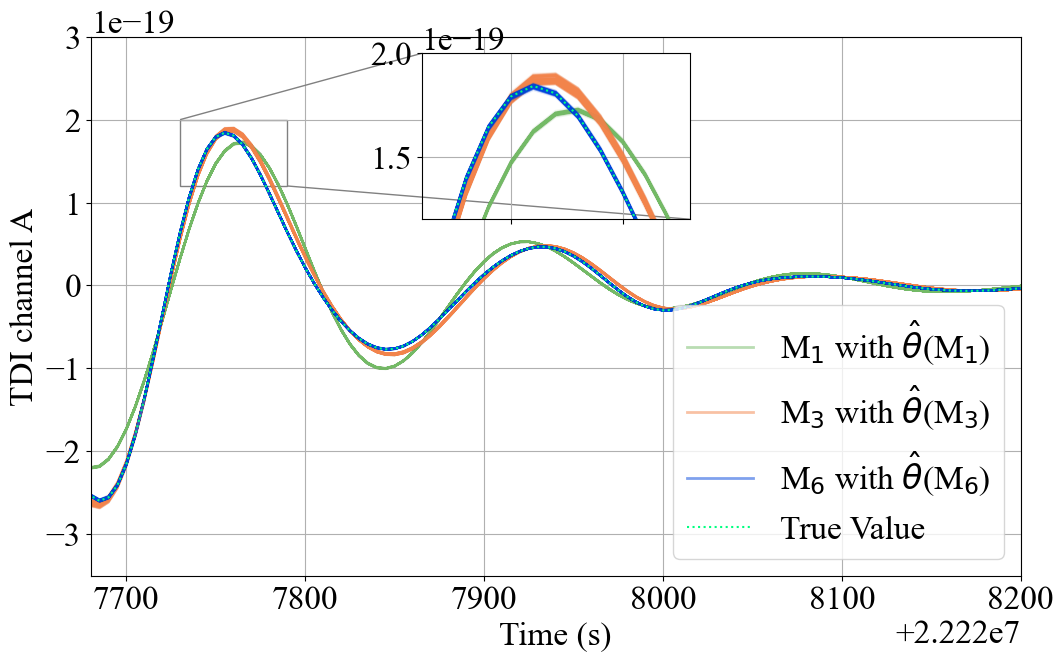}
    \hypertarget{fig:amphase_models}{}
    \caption*{a) Waveform obtained with models $M_1$, $M_3$, and $M_6$ with parameters sampled from posterior distributions of the corresponding model.}
    \includegraphics[width=0.45\textwidth]{ 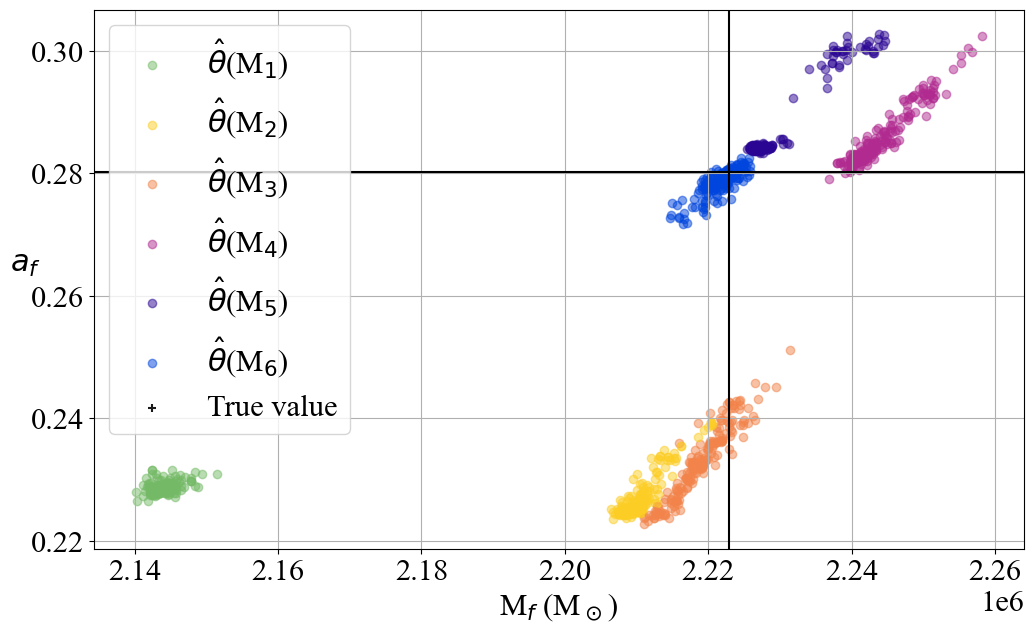}
    \hypertarget{fig:spin_mass}{}
    \caption*{b) Derivation of final BH mass and spin from posteriors using models $M_1$ to $M_6$ compared with the true values.}
    \caption{\label{fig:ringdowns}Illustrative effect of biased parameters in the ringdown. In the top figure, we see waveforms generated with the same model as the one used in the inference, obtained from 200 posterior samples for three models $\hat{\theta}$($M_{1,3,6}$). We can see the consistency of the models with the injection, albeit its parameter bias. The bottom figure shows the mass and spin of the remnant BH derived from each set of parameters for all models (colored dots) and the true value (crossing black lines). This is a visual representation of the impact of biased parameters on the remnant BH.}
\end{figure}

In Fig.~\ref{fig:ringdowns}, we illustrate the impact of the parameter biases on the reconstruction of the post-merger waveform. We randomly select 200 samples from posterior distributions obtained with models $M_1$, $M_3$, and $M_6$ and generate the waveform in the time domain with the same model. Fig.~\hyperlink{fig:amphase_models}{\ref{fig:ringdowns}a} serves as a visual representation of the amplitude and phase from the results obtained by each model. We cannot distinguish individual lines due to the small statistical error: the posteriors are centered around biased parameters but with a small dispersion. The waveform reconstruction would therefore be ``confidently wrong". Trying to infer a ringdown analysis with IMR information from biased analyses would presumably raise issues. If we compare $M_3$ and $M_6$ (in orange and blue, respectively), enlarging the image, we see visible differences in the post-merger phase. This feature, consistently with the significant Bayes factor for model $M_6$ over model $M_3$, highlights the contribution of less dominant modes such as (2,1), (3,2), and (4,3). 

One of the tests looking for deviations from GR in ringdown signals consists in comparing the final mass and spin inferred from the ringdown signal with the values derived from the IMR posteriors using formulas fitted on numerical relativity; the consistency between the two estimates is the focus of the test. We do not perform a ringdown analysis here. Still, we illustrate in Fig.~\hypertarget{fig:mass_spin}{\ref{fig:ringdowns}b} how the parameter biases found in our IMR parameter estimation would translate into erroneous mass and spin. Using the same fitted formulas as in PhenomHM~\cite{London_2018} (see Eqs.~(3.6) to (3.8) in \cite{PhenomD}), we derived the final mass and spin for 2000 randomly distributed points within the posterior distribution for each model. The IMR parameter biases would translate into significantly biased $M_f$ and $a_f$. In this figure, the addition of higher modes shows no clear trend for the final mass, but we can appreciate how the introduction of weaker higher modes helps to obtain a more accurate final spin for the remnant BH. Those models are $M_4$, $M_5$, and $ M_6$, in pink, purple, and blue, respectively. It is worth mentioning the scattered distribution of $M_5$, which is the result of the bimodal distribution of individual spins (see Fig.\ref{fig:res}). 


\subsection{Modelling error and SNR dependency}

\begin{figure*}[t]
    \centering
    \includegraphics[width=0.95\textwidth]{ 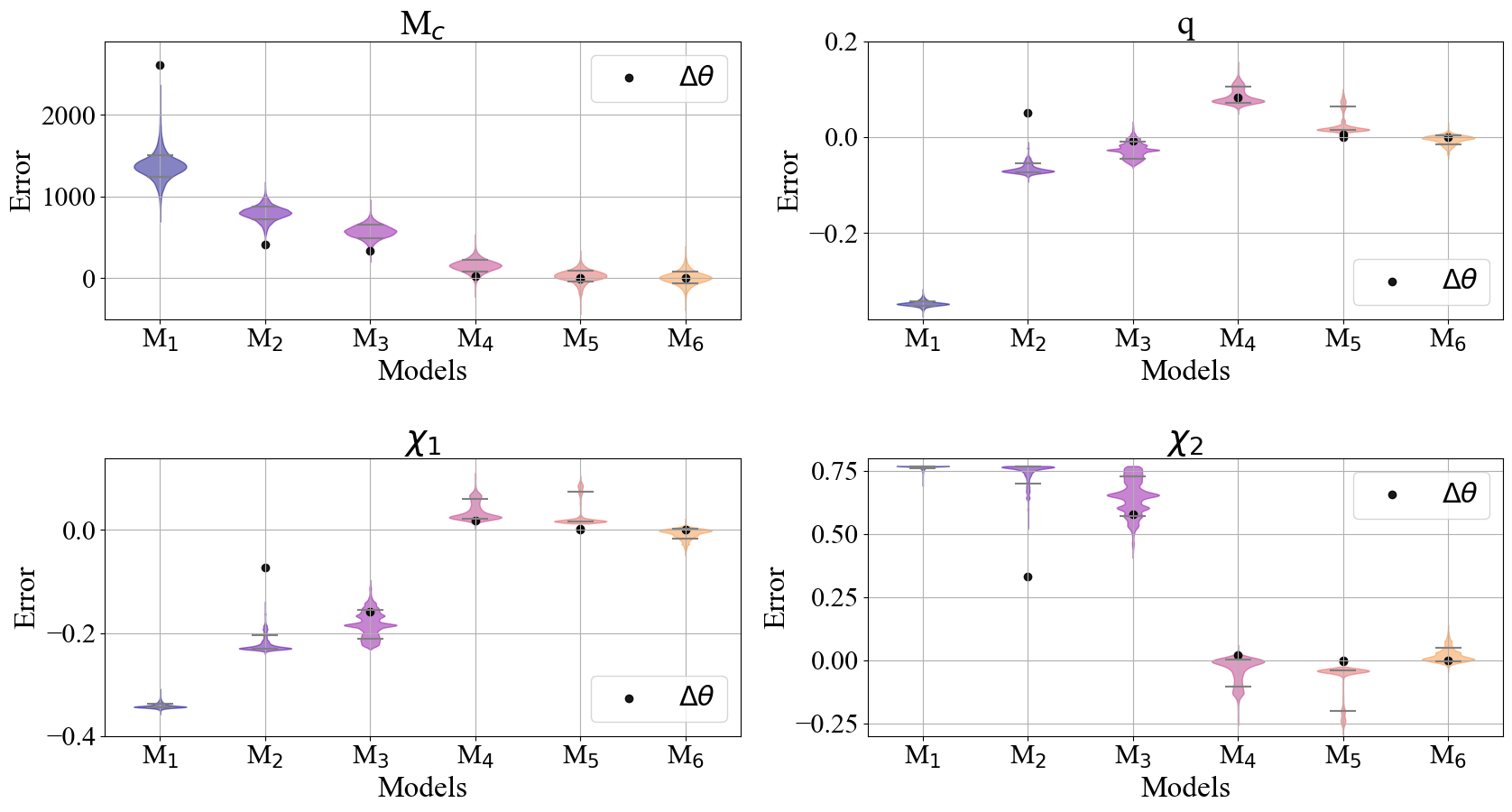}
    \caption{Comparison of the error in the intrinsic parameters posterior distribution for all models ($M_k$, $k=1,\hdots,6$ in color) with the error derived from the Fisher approximation (in black). Due to its large value, the Fisher bias for $M_1$ in the spins and mass ratio lies outside the plot. We observe consistency between the Fisher computation and posteriors for models $M_k$ with $k=3,4,5,6$ as the bias decreases.}
    \label{fig:violin_deltatheta}
\end{figure*}

\begin{figure*}[t]
    \centering
    \includegraphics[width=0.9\textwidth]{ 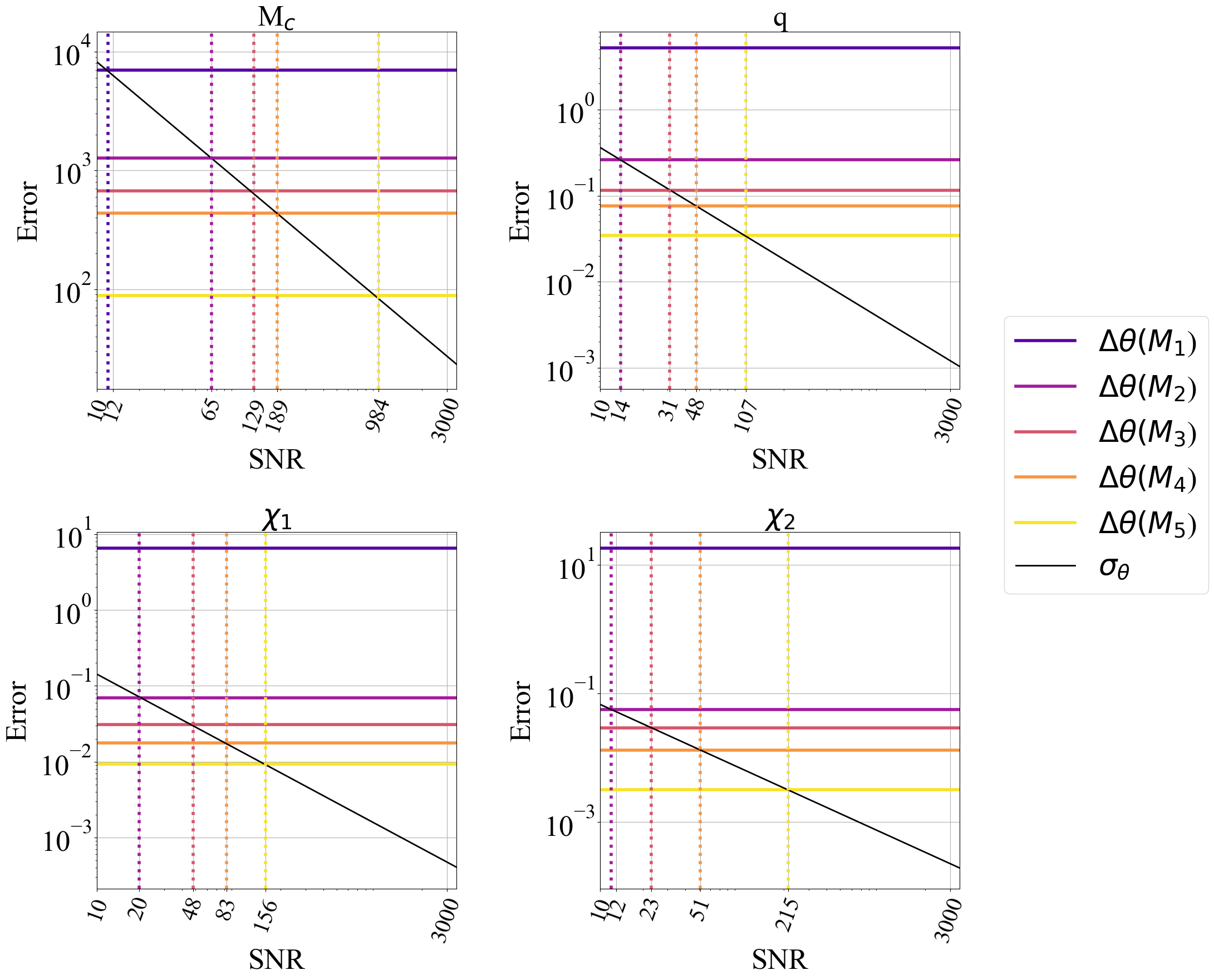}
    \caption{Comparison of the statistical error (in black) and modeling errors (in colors) for relevant parameters in function of the SNR. When the statistical error becomes smaller than the modeling error, that model no longer fits the data well and biases the estimated parameters. The SNR for this transition is marked with dotted lines in the correspondent color for each incorrect template (models $M_k$, $k=1,\hdots,5$).}
    \label{fig:deltatheta}
\end{figure*}

The magnitude of the bias on each parameter depends on the SNR and on the template waveform used for the inference, hence on the model $M_k$ (with $k=1,\hdots,6$). To properly analyze this issue, we introduce some definitions. Within the Fisher approximation, valid in principle in the high-SNR limit, the statistical error in each parameter $\sigma_{\theta}$ produced by the noise for a given waveform is determined as
\begin{equation}\label{eq_sigmatheta}
    \sigma_{\theta_i}=\sqrt{\Gamma^{-1}_{ii}},
\end{equation}
where $\Gamma_{ij} = \left( \pdv{h}{\theta_i} \vert \pdv{h}{\theta_j}\right)$ is the Fisher information matrix. In this framework, the statistical error scales directly as $\mathrm{SNR}^{-1}$. 

The bias $\Delta \theta(M_k)$ or ``modeling error" due to the use of an incorrect template is defined as~\cite{Cutler_2007}
\begin{equation}\label{deltatheta}
    \Delta \theta_i =\hat{\theta}_i^0 - \hat{\theta}_i^{\text{temp}}= \sum_j \Gamma_{ij}^{-1}(\theta_{k}) \, \left ( \pdv{h_k}{\theta_j} \vert \delta h_k \right),
\end{equation}
where $k$ refers to model $M_k$, with $k = 1,\, \hdots,\, 5$ and $\delta h_k = h_{0} - h_k$ is the difference between the true waveform and the template waveform with model $M_k$. The template model corresponds to waveforms generated with the modes defined in Table~\ref{tab:models}, so $\delta h_k$ is simply the sum of the ignored modes.

If the statistical error $\sigma_{\theta}$ is larger than the error produced using an incorrect template $\Delta \theta (M_k)$, one can consider the bias irrelevant. On the contrary, if the statistical error is smaller than the modeling error, the waveform model is not sufficiently accurate to describe the data, and the bias becomes relevant.

We first check in Fig.~\ref{fig:violin_deltatheta} whether the bias observed in the posterior distributions is consistent with the value obtained from the approximate Fisher bias formula in Eq.\eqref{deltatheta}. We show the bias from each model (black dot) and the error distribution obtained from the sampler (in colors) for intrinsic parameters. In the case of $M_1$, we observe that the Fisher bias is much larger than the one found by sampling for all parameters, sometimes lying outside the plot. For adimensional spins, this may be because their values are limited to the interval [-1, 1] in the sampler, whereas they are unconstrained in the Fisher matrix computation. In the case of $M_2$, the opposite happens, and the posterior distribution exhibits a slightly larger error than the bias predicted by the Fisher analysis. Overall, from model $M_3$ to $M_6$, both errors become more and more consistent as parameter biases shrink. Thus, we can rely on the analytical bias obtained with the Fisher approximation for the intrinsic parameters. The same analysis for extrinsic parameters gave disagreeing results, with typically an overestimation of the bias with the Fisher approximation compared to the sampled posteriors; see Appendix~\ref{app_1} for a short discussion on this matter. 

A natural question arises about the minimum SNR at which higher modes become important in parameter estimation. In other words, given a certain SNR, how many modes do we need to describe the waveform adequately? One way to answer this question is by comparing the approximate statistical error for each parameter $\sigma_{\theta}$ with the systematic bias induced by an incorrect model $\Delta \theta(M_k)$. We perform this comparison in Fig.~\ref{fig:deltatheta}, where we show the errors for the intrinsic parameters as a function of SNR. Varying the SNR amounts to changing the value of the luminosity distance $D_L$, leaving all other parameters unchanged. The Fisher bias in models $M_1$ and $M_2$ are inconsistent with posteriors, as observed in Fig.\ref{fig:violin_deltatheta}.
For this reason, we will not discuss them, although they are plotted in the figure. The black diagonal line corresponds to the statistical error in the model parameters ($\sigma_{\theta}$), and the color lines represent the modeling error produced by the wrong waveform template ($M_k$). With the same color code as the modeling error, we mark the value of SNR in dotted lines at which the modeling error becomes higher than the statistical error.

Fig.~\ref{fig:deltatheta}'s top left panel shows that the model $M_3$ (in pink), which includes modes (2, 2), (3, 3), and (4, 4), does not describe accurately enough the signal for sources with SNR $\geq 129$ since the chirp mass bias becomes more significant than the statistical error. Similarly, the model $M_4$ is sufficient until the SNR reaches 189 and $M_5$ until 984. Since the SNR of the source we consider in this work is around 744, the value of the chirp mass inferred with the model $M_5$ (in yellow) should be within the estimated error. However, to accurately infer the other parameters, the model $M_5$ works until an SNR of 107 for the mass ratio, 156 for $\chi_1$, and 215 for $\chi_2$. Then, with an SNR of 744, this model will correctly estimate the chirp mass but induce biases for all other parameters. For consistency, we can look at the inferred values from the Bayesian analysis with $M_5$ in Fig.~\ref{fig:res} and confirm this statement within the 68\% confidence level.

This analysis does not derive a limit on the number of modes needed to describe an event observed by LISA. Still, it provides maximum SNR values to correctly estimate the parameters with a given model if less than these six modes are present in the waveform. Extrapolating from our example, we find that generally, sources detected by LISA with SNR of hundreds will require using waveforms with at least six modes to estimate all intrinsic parameters correctly. Note that the Fisher error in Eq.\eqref{eq_sigmatheta} and its scaling with $\mathrm{SNR}^{-1}$ are only approximate and are best valid for high-SNR and non-degenerate posteriors, so this estimate does not replace a complete parameter estimation study. 

\section{Conclusions}\label{sec6}

We studied the contribution of IMR higher modes of a SMBHB source to the total SNR. We observed how this contribution depends on the event's redshifted mass through the observed frequency, directly related to LISA's response and sensitivity. We also showed that the cross-terms could contribute constructively or destructively to the total SNR, depending on the signal frequency and observational parameters. We presented a map guide of the relevance of each mode given the mass of an event. We highlighted the role of higher modes for SMBHBs with masses of the order of $10^8$ M$_\odot$. In LISA, large mass sources enhance the contribution of modes with higher frequencies so that the quadrupolar mode will no longer dominate. 

To compare sensitivity performances, we defined different models, each including different harmonics. We could distinguish higher modes by comparing the Bayesian evidence for different models. In our example of a noisy signal with six modes, the model $M_6$, which includes the same higher modes as in the injection, was the preferred one, as expected. The model $M_6$ showed a very significant Bayes factor compared to models with fewer modes. Furthermore, we found that the absence of modes in the waveform template can bias the parameter estimation for high SNR sources due to the non-orthogonality of the modes in the merger-ringdown phase. Biased binary parameters can lead to a biased inference of the remnant BH's mass and spin. This effect can corrupt the no-hair theorem test and lead to misinterpretations.

We could quantify the SNR needed to distinguish models by comparing statistical errors of the injected waveform parameters with the modeling errors produced using an incorrect template ($M_k$, $k=1,\hdots,5$). In other words, given a certain SNR, we can specify the modes needed to infer the parameters without significant bias. This quantification depends on the actual waveform, which includes six modes in our analysis. In reality, such a situation is unlikely, as we expect more modes in the dynamics. Hence, this study does not derive a limit on the number of observable modes, which is still an open question that can be answered once more harmonics are implemented. However, our work demonstrates the need for higher modes in the waveform templates to perform accurate GW source characterization with LISA. Besides, featuring precession and eccentricity in the inspiral stage will also be necessary, while mode-mixing and non-linearity \cite{London_2014, Berti_2014, Mark_22, Mitman_22} will become essential features in the ringdown.

The ability of LISA to identify different modes allows us to consider GR tests on more solid grounds, including the test of the no-hair theorem, which will be the subject of a forthcoming study.


\begin{acknowledgments}

This work was supported by CNES, focused on LISA Mission. We gratefully acknowledge support from the CNRS/IN2P3 Computing Center (Lyon - France) for providing computing and data-processing resources needed for this work. CP acknowledges financial support from CNES Grant No. 51/20082 and CEA Grant No. 2022-039.
\end{acknowledgments}
\begin{appendix}

\section{Complementary results}\label{secA2}

\begin{figure*}
    \centering
    \includegraphics[width=0.7\textwidth]{ 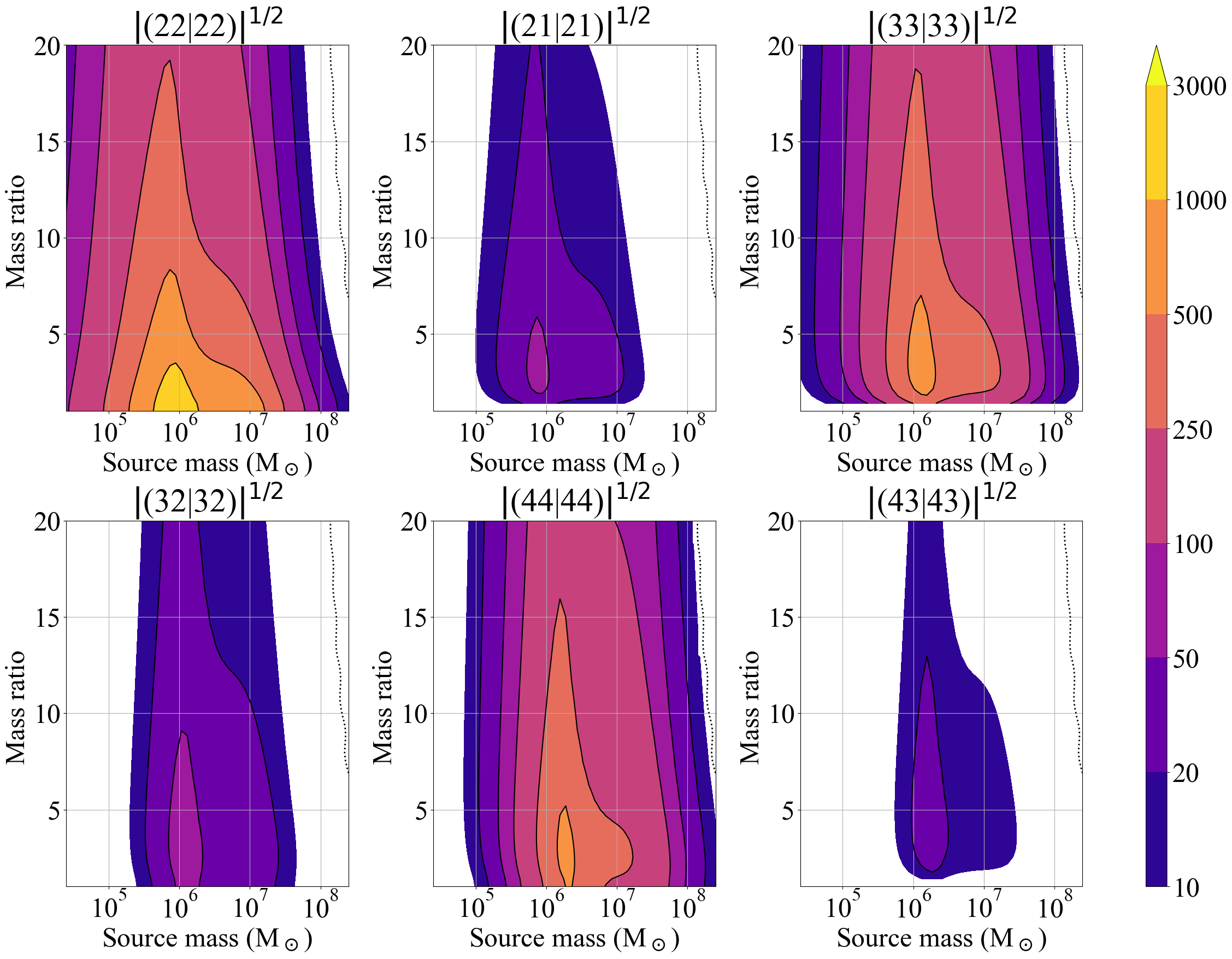}
    \caption{Contour plots for mass ratio with respect to the total mass. The plot highlights the contribution of higher modes for sources of $\sim$ $10^6$ M$_\odot$ and mass ratios up to 15. SNR decreases slowly when increasing the mass ratio by a few units for a fixed mass.}
    \label{fig:waterfall_q_sym}
\end{figure*}

To assess how the mass ratio affects the SNR, we show in Fig.~\ref{fig:waterfall_q_sym} the contour plots representing mass ratio versus total source mass for a source with parameters given in Table~\ref{tab:params} and redshift $z=2$. The absence of contribution at leading PN order of modes with odd $m$ for equal mass ratio (see e.g.\cite{Arun_08}) is illustrated here. We can also note the importance of higher modes for sources with masses around $10^6 \, M_\odot$, and mass ratios between 2 and 15.

For completeness, we also include here the square and cross-terms of the SNR contributions from Section~\ref{sec_mass}. As well as the marginalized posterior on all parameters for $M_1$, $M_3$, and $M_6$ with noisy data from Section~\ref{sec_results}.

\begin{figure*}
    \centering
    \includegraphics[width = 0.88\textwidth ]{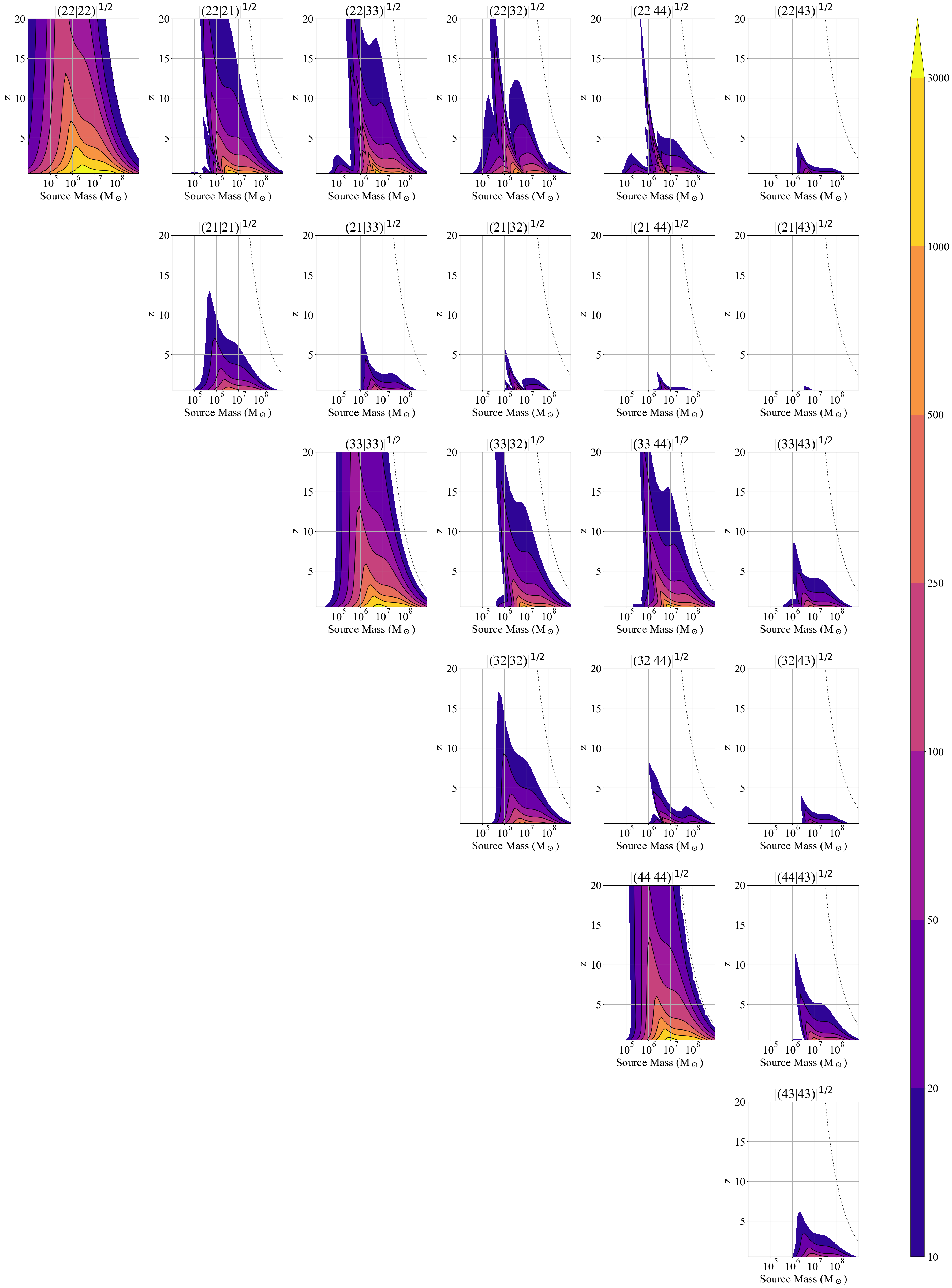}
    \caption{SNR contribution for each pair of modes}
    \label{fig:waterfall_all}
\end{figure*}

\begin{figure*}
    \centering
    \includegraphics[width = \textwidth ]{ 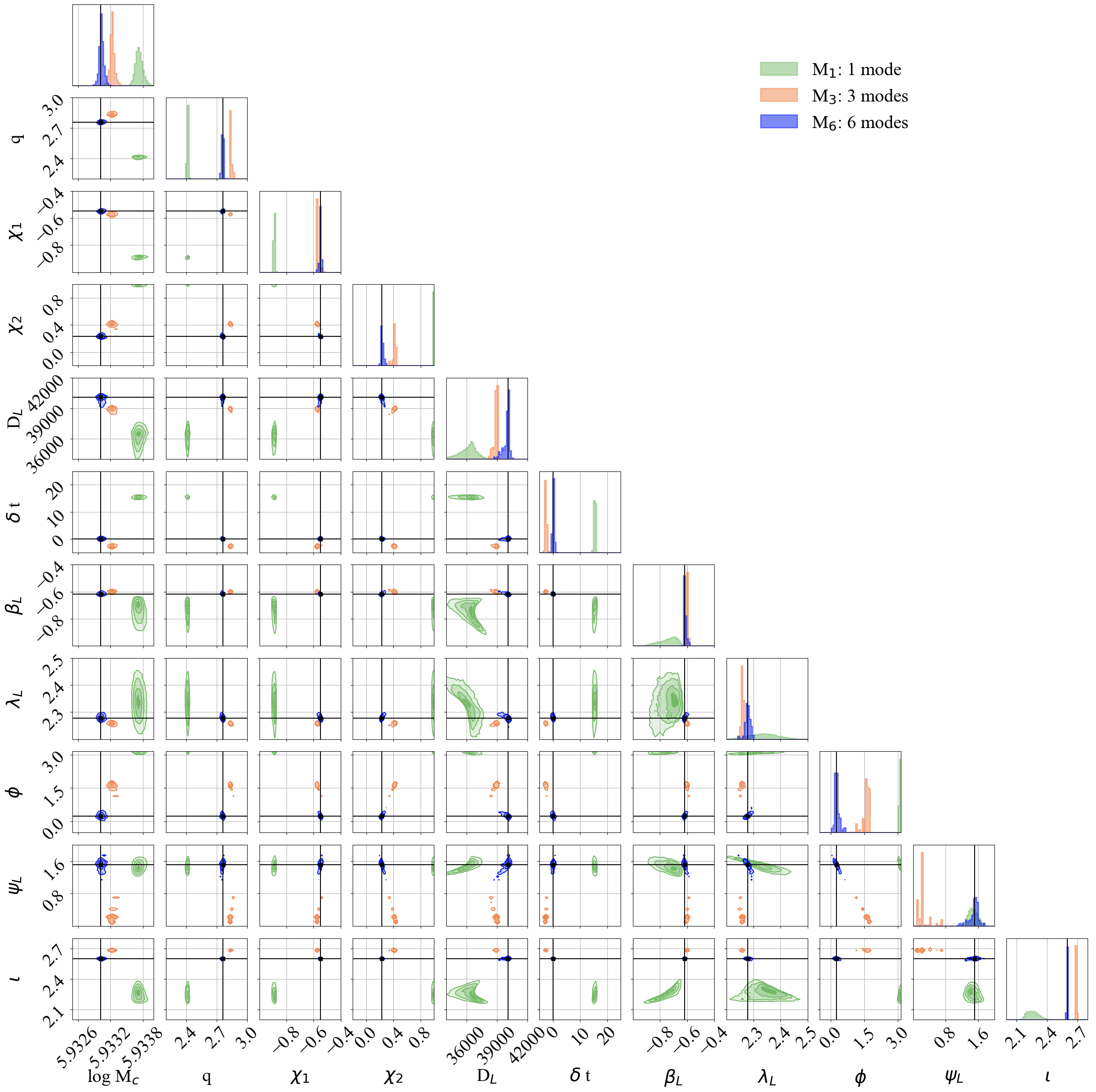}
    \caption{Marginalised posterior distribution for all parameters. Here $M_1$ in green, $M_3$ in orange and $M_6$ in blue with true values represented with black lines. Note the high accuracy of the estimated mean value of all  parameters found with the model $M_6$.}
    \label{fig:res_noise_all}
\end{figure*}

\subsection{Modelling error}\label{app_1}
\begin{figure*}
    \centering
    \includegraphics[width = \textwidth]{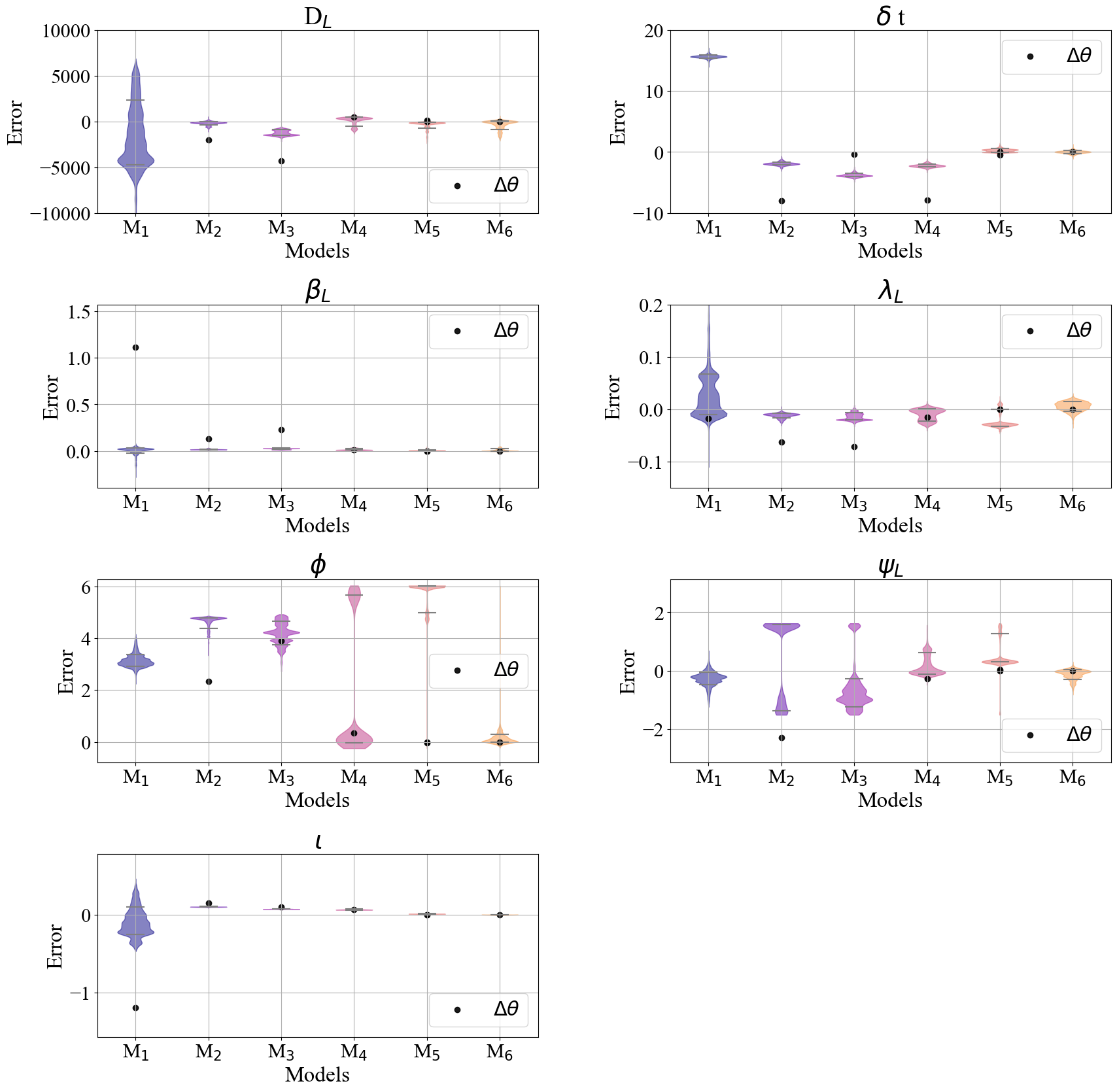}
    \caption{Comparison of the error in the posterior distribution of extrinsic parameters for all models ($M_k$, $k=1,\hdots,6$ in color) and modelling error (in black).}
    \label{fig:violin_ex}
\end{figure*}

If we compare the modeling error on the extrinsic parameters obtained from the Fisher information with the ones obtained from posterior distributions, see Fig.\ref{fig:violin_ex}, we observe some disagreeing results. The model $M_1$ is the least accurate, as some Fisher bias values lie outside the range in the plots. As we introduce more modes, the difference between analytical and experimental results tends toward zero. Thus, we expect the analytical bias to be trustworthy for $M_k$ with $k\geq 4$. As previously mentioned, the discrepancy could come from the constraint set on the parameter space explored by the sampler, which is absent from the Fisher derivation. This can be seen for example in the polarization $\psi_L$ and phase $\phi$, where some analytical points are outside the allowed range. Another possible explanation is the multimodality of some extrinsic parameters, such as the ecliptic latitude $\beta_L$ in LISA's frame and the source inclination $\iota$, especially for $M_1$. We also observe that most of the errors obtained with the Fisher information can change if we use for instance adimensional spin parameters such as $\chi_{+}$ and $\chi_{-}$ instead of individual spins $\chi_{1}$ and $\chi_{2}$. Furthermore, the numerics of Fisher matrices are notoriously delicate, and we leave the observed discrepancy between extrinsic errors for future investigation.

\end{appendix}
\bibliography{new_version}

\end{document}